\documentclass[a4paper,11pt]{article}
\pdfoutput=1

\usepackage{jheppub}

\usepackage[T1]{fontenc}
\usepackage{amssymb,amsmath}
\usepackage{braket,bbm,bm}
\usepackage{mathtools}
\usepackage[usenames,dvipsnames,svgnames,table]{xcolor}
\usepackage[utf8]{inputenc}
\usepackage{soul, color}
\usepackage{subcaption}
\usepackage{lmodern}
\usepackage{footnote}
\usepackage[normalem]{ulem}
\usepackage{glossaries-extra}
\setabbreviationstyle[acronym]{long-short}
\glssetcategoryattribute{acronym}{nohyperfirst}{true}

\usepackage{slashed}
\usepackage{multirow}
\usepackage{here}
\usepackage{hyperref}
\usepackage{verbatim}
\usepackage[justification=justified,singlelinecheck=false]{caption}
\usepackage{cleveref}
\usepackage{listings}
\usepackage{fancyvrb}
\usepackage{dsfont}
\usepackage{nicefrac,xfrac}
\usepackage{verbatim}

\Crefname{equation}{eq.}{eqs.}
\Crefname{section}{section}{sections}
\Crefname{figure}{figure}{figures}
\Crefname{appendix}{appendix}{appendices}

\newcommand{\cA}{\mathcal{A}}
\newcommand{\cB}{\mathcal{B}}
\newcommand{\cC}{\mathcal{C}}
\newcommand{\cG}{\mathcal{G}}
\newcommand{\cE}{\mathcal{E}}
\newcommand{\cD}[0]{\mathcal D}

\newcommand{\cI}[0]{\mathcal I}

\newcommand{\cK}[0]{\mathcal K}
\newcommand{\cL}[0]{\mathcal L}
\newcommand{\cM}[0]{\mathcal M}
\newcommand{\cO}[0]{\mathcal O}
\newcommand{\cR}[0]{\mathcal R}

\newcommand{\cT}[0]{\mathcal T}

\usepackage{mfirstuc} 
\newcommand{\addReviewer}[2]{
  \expandafter\newcommand\csname #1\endcsname[1]{{\bf \color{#2} \capitalisewords{#1}:\,##1}}
  \expandafter\newcommand\csname #1cor\endcsname[2]{{\color{#2} \capitalisewords{#1}:\,\st{##1}{\bf ##2}}}
  \expandafter\newcommand\csname #1color\endcsname{#2}
}

\definecolor{cardinal}{rgb}{0.77, 0.12, 0.23}

\addReviewer{sebastian}{cardinal}
\addReviewer{fernando}{orange}
\addReviewer{steve}{Maroon}

\newcommand{\HSQCa}[0]{Hansen:2014eka}
\newcommand{\HSQCb}[0]{Hansen:2015zga}

\newcommand{\BHSnum}[0]{Briceno:2018mlh}

\newcommand{\tetraquark}[0]{Hansen:2024ffk}
\newcommand{\BSQC}[0]{Blanton:2020jnm}

\newcommand{\BSnondegen}[0]{Blanton:2020gmf}
\newcommand{\BStwoplusone}[0]{Blanton:2021mih}
\newcommand{\DRS}[0]{Dawid:2024dgy}
\title{\boldmath 
Comparison of integral equations used to study $T_{cc}^+$ for a stable $D^*$
}

\author[a]{Sebastian M. Dawid}
\author[b]{, Fernando Romero-L\'opez}
\author[a]{, and Stephen R. Sharpe}

\affiliation[a]{Physics Department, University of Washington, Seattle, WA, 98195-1560, USA}
\affiliation[b]{Albert Einstein Center, Institute for Theoretical Physics, University of Bern, 3012 Bern, Switzerland}

\emailAdd{dawids@uw.edu}
\emailAdd{fernando.romero-lopez@unibe.ch}
\emailAdd{srsharpe@uw.edu}

\abstract{
We perform a detailed comparison between three formalisms used in recent studies of $DD^*$ scattering at heavier-than-physical pion masses, which aim to understand the properties of the doubly-charmed tetraquark, $T_{cc}^+(3875)$.
These methods are the three-particle relativistic field theory (RFT) formalism, the two-body Lippmann–Schwinger (LS) equation with chiral effective field theory potentials, and the two-particle relativistic framework proposed by Bai\~ao Raposo and Hansen (BRH approach). In a simplified single-channel setting, we derive the conditions under which the infinite-volume integral equations from the RFT and BRH approaches reduce to the LS form. We present numerical examples showing that differences between these methods can be largely removed by adjusting short-range couplings. We also address a number of technical issues in the RFT approach.
}
\allowdisplaybreaks

\begin{document} 
\maketitle
\clearpage
\flushbottom

\section{Introduction}
\label{sec:intro}

The discovery of the doubly-charmed tetraquark resonance $T_{\rm cc}^+(3875)$~\cite{LHCb:2021vvq, LHCb:2021auc} has led to intense theoretical activity aiming to understand its properties from the underlying theory, QCD~\cite{Meng:2021uhz,Du:2021zzh, Meng:2022ozq, Wang:2023iaz, Zhai:2023ejo,Du:2023hlu, Raposo:2023oru, Meng:2023bmz, Bubna:2024izx, Collins:2024sfi, Dawid:2024dgy,\tetraquark,Meng:2024kkp, Sun:2024wxz,Zhang:2024dth, Abolnikov:2024key,  Whyte:2024ihh,  Stump:2024lqx, Prelovsek:2025vbr,Raposo:2025dkb, Lachini:2025lwd}. This exotic hadron is observed experimentally as a resonance decaying to a three-body $DD\pi$ final state. By contrast, lattice QCD computations performed at heavier-than-physical quark masses---for which the $D^*$ meson is stable---reveal it as a subthreshold complex pole in the $DD^*$ scattering amplitude~\cite{Du:2023hlu, Collins:2024sfi, Prelovsek:2025vbr}. A robust theoretical description of the $T_{\rm cc}^+$ requires addressing its three-body decay at the physical point, and the effects of one-pion exchange (OPE) processes at heavier quark masses. The latter, in particular, implies a non-analyticity---a left-hand cut (LHC)---in the $DD^*$ scattering amplitude that lies close to the location of the $T_{\rm cc}^+$ pole and thus cannot be ignored.

A common approach that includes the correct subthreshold analytic structure of the $DD^*$ amplitude uses the Lippmann--Schwinger (LS) equation. In this approach, the $D$ and $D^*$ are treated nonrelativistically, while one-pion exchange is included as a potential arising from chiral effective field theory (EFT), in which the pion kinematics can be treated relativistically~\cite{Du:2021zzh, Meng:2022ozq, Wang:2023iaz, Zhai:2023ejo, Meng:2023bmz, Collins:2024sfi, Meng:2024kkp, Sun:2024wxz, Abolnikov:2024key, Prelovsek:2025vbr}.\footnote{%
This approach is based on an earlier study of the closely related $\chi_{c1}(3872)$ state~\cite{Baru:2011rs}.}
This method has been applied both to the analysis of lattice QCD results from simulations with heavier-than-physical quarks, for which the $D^*$ is stable, and to the investigation of the physical $T_{\rm cc}^+$, using various model parameters to describe the interactions of the unstable $D^*$ and the $D$~\cite{Du:2021zzh}.\footnote{See also~\cite{Zhang:2024dth} for a closely related treatment.}

An alternative theoretical approach has recently been proposed~\cite{\tetraquark}, and subsequently implemented~\cite{Dawid:2024dgy}, in which the three-particle $DD\pi$ system is treated fully relativistically. This formalism contains both a finite-volume part (for fitting lattice QCD results) and an infinite-volume part. Only the latter is relevant for the present work---it provides a set of infinite-volume integral equations for the elastic $DD\pi$ scattering amplitude. The $D^*$ enters this formalism as either a bound state or a resonance in the $D\pi$ subchannel, and this approach allows for a natural interpolation between these scenarios (for instance, as a function of varying pion mass). For a stable $D^*$, the $DD^*$ amplitude is obtained from the $DD\pi$ amplitude via Lehmann-Symanzik-Zimmermann (LSZ) reduction, i.e., by considering subthreshold, $p$-wave, $D\pi$ subsystems, and determining the residue of the three-body amplitude at the $D^*$ pole. In a nutshell, this approach studies three-body interactions of the $DD\pi$ system, with the two-particle amplitude obtained as a derived quantity.

In the field of amplitude analysis, the three-body RFT approach is relatively new~\cite{Jackura:2020bsk, Jackura:2022gib, Dawid:2023jrj, Dawid:2023kxu, Jackura:2023qtp,  Briceno:2024ehy, Dawid:2025zxc, Dawid:2025doq} 
\footnote{%
Infinite-volume three-body amplitudes were also studied recently using the equivalent finite-volume unitarity (FVU) method~\cite{Mai:2017vot, Jackura:2018xnx, Dawid:2020uhn, Sadasivan:2020syi, Sadasivan:2021emk, Feng:2024wyg,Sakthivasan:2024uwd}.} 
This approach originates from the three-particle formalism of the generic relativistic field theory (RFT) method for extracting scattering amplitudes (and thus resonance parameters) from finite-volume spectra~\cite{\HSQCa,\HSQCb, Briceno:2018aml}. In the spirit of S-matrix theory, the $3 \to 3$ amplitudes are described by an equation in which no assumptions are made about their microscopic origin, but that ensure consistency with the fundamental physical properties, such as S-matrix unitarity~\cite{Jackura:2019bmu, Briceno:2019muc, Jackura:2022gib}. In the RFT approach, these amplitudes are described by on-shell integral equations taking the two- and three-particle $K$ matrices as input. These are meromorphic functions of the kinematic variables that, in practice, are parametrized solely based on their desired analytic properties. 

The LS approach, which has a long history,~\cite{Lippmann:1950zz, Gell-Mann:1953dcn} is, {\em prima facie}, quite different. Apart from its largely nonrelativistic context, it involves off-shell quantities, like potentials, and considers only two particles (here, the $DD^*$ system). Thus, the exact nature of its connection with the three-body RFT approach is not immediately apparent. Establishing such a connection is the main goal of this work.

One motivation for the proposal to study the $D D^*$ system using the three-particle approach is that it automatically accounts for the finite-volume effects originating from the LHC associated with $u$-channel OPE~\cite{\tetraquark}. In contrast, the standard finite-volume two-particle L\"uscher approach~\cite{Luscher:1986n1, Luscher:1986n2, Rummukainen:1995vs, Kim:2005gf} breaks down below the opening of this cut and cannot be used in the vicinity of the $T_{\rm cc}^+$.%
\footnote{%
This issue was first noted in two-nucleon systems~\cite{Green:2021qol}.
}
Other methods to resolve this ``LHC problem'' have been proposed~\cite{Klos:2018sen, Meng:2021uhz, Raposo:2023oru, Bubna:2024izx, Dawid:2024oey}. One, by Meng and Epelbaum, is based on implementing the above-mentioned chiral EFT in finite-volume in the plane-wave basis~\cite{Meng:2021uhz}; in this case, the associated infinite-volume equation is in the LS class~\cite{Meng:2023bmz, Meng:2024kkp, Abolnikov:2024key, Prelovsek:2025vbr}. Another proposal, by Bai\~ao~Raposo and Hansen~\cite{Raposo:2023oru}, subsequently generalized by Bai\~ao Raposo, Brice\~no, Hansen, and Jackura in ref.~\cite{Raposo:2025dkb}, amends the two-particle formalism by singling out OPE diagrams and treating them explicitly. It leads to a generalization of the L\"uscher approach that describes the $DD^*$ system by a relativistic two-particle integral equation, the ``BRH equation'', that differs from the nonrelativistic LS equation.

We are thus in a situation where there exist (at least) three approaches available to study the $D D^*$ (and related) systems,  both in finite and infinite volume. It is clearly of interest to determine the relation between them. We begin the task in this article by comparing the infinite-volume equations of the different formalisms, focusing on the partially nonrelativistic (NR) regime in which the LS approach is formulated. For the sake of simplicity and brevity, we make the comparison in the simplified setting in which only interactions in the dominant partial wave are included.

This paper is organized as follows. In \Cref{sec:RFT}, we briefly describe the general form of the three-particle integral equations of the RFT approach, and then present, in full detail, their simplified limit in our single-channel approximation. We also show how, using the LSZ reduction, one obtains a two-particle equation. In~\Cref{sec:LSE}, we review the LS equation used in ref.~\cite{Collins:2024sfi} and present its single-channel version. In~\Cref{sec:NR-RFT}, we describe the approximation in which the RFT integral equations match those of the LS approach: a combined NR limit and pole-dominance approximation. In \Cref{sec:numerical}, we provide a numerical demonstration of the relation between the RFT, LS, and BRH approaches for parameters that are close to those describing the lattice results of ref.~\cite{Padmanath:2022cvl}. We summarize and conclude in \Cref{sec:conclusions}.

We include three appendices. In \Cref{app:RH} we briefly summarize the BRH equations, present their form in the one-channel $DD^*$ system, consider their NR limit, and compare with the other approaches. In  \Cref{app:cutoff} we describe technical issues concerning the cutoff used in the RFT approach. Finally, in~\Cref{app:offshell}, we study the impact of relaxing the pole-dominance approximation on the matching between RFT and LS approaches.

\section{Three-particle RFT approach}
\label{sec:RFT}

In the approach of refs.~\cite{\tetraquark,\DRS}, $D D^*$ scattering for a stable $D^*$ and in the isospin limit is studied by considering the elastic ${D D \pi \to D D \pi}$ process in the isospin $I=0$ channel. The $DD^*$ amplitude is then obtained via the LSZ reduction formula.

\subsection{General formalism}
\label{sec:RFT:gen}

To flesh out the above words, we first summarize the integral equations satisfied by the on-shell $3\to3$ amplitude. We use the compact notation of ref.~\cite{\DRS}, to which we direct the reader for a more complete description. The relevant parts of the notation will be explained as needed. The integral equations in the RFT approach were first derived for identical scalars in ref.~\cite{\HSQCb}, with the generalization to the $DD\pi$ system given in ref.~\cite{\tetraquark}, extending the work of refs.~\cite{\BSnondegen, \BStwoplusone}.
The $DD\pi$ amplitude is decomposed as
    \begin{align}
    \bm{\cM}_3 = \bm{\cD} + \bm{\cM}_{\text{df},3} \,,
    \label{eq:M3}
    \end{align}
where the ladder amplitude, $\bm{\cD}$, satisfies the integral equation
    \begin{align}
    \bm{\cD}= - \bm{\cM}_2 \, \bm{G} \, \bm{\cM}_2 - \bm{\cM}_2 \, \bm{G} \, \bm{\cD} \,,
    \label{eq:D}
    \end{align}
which sums up contributions from one-particle exchanges (entering through the OPE kernel $\bm G$) between two-particle scatterings (described by the $2\to2$ amplitudes $\bm{\cM}_2$). To apply the LSZ reduction, it is useful to define an ``amputated'' amplitude through
    \begin{align}
    \bm{d} = \bm{\cM}_2^{-1} \, \bm{\cD} \, \bm{\cM}_2^{-1} \, ,
    \label{eq:dfromD}
    \end{align}
which satisfies the modified integral equation
    \begin{align}
    \bm{d}= - \bm{G} - \bm{G} \, \bm{\cM}_2 \, \bm{d} \,.
    \label{eq:inteqd}
    \end{align}
This form of the ladder integral equation was solved numerically in ref.~\cite{\DRS}.

The second term in \Cref{eq:M3} includes the effect of short-range, three-particle interactions, described by a three-particle $K$ matrix, $\bm{\cK}_3$. It is given by
    \begin{align}
    \bm{\cM}_{\text{df},3} = \bm{\cL} \, \bm{\cT} \, \bm{\cR} \, ,
    \label{eq:LTR}
    \end{align}
where the ``endcaps'' $\bm{\cL}$ and $\bm{\cR}$ incorporate final-state two-particle interactions and OPEs. $\bm{\cT}$ satisfies
    \begin{align}
    \bm{\cT} = \bm{\cK}_3 - \bm{\cK}_3 \, \bm{\tilde{\rho}} \, \bm{\cL} \, \bm{\cT} \, ,
    \label{eq:T}
    \end{align}
where $\bm{\tilde{\rho}}$ is a phase-space operator.
The subscript ``df'' on $ \bm{\cM}_{\text{df},3}$
stands for ``divergence-free,'' which indicates that all potential on-shell singularities associated with OPE in the three-particle amplitude are contained in $\bm{\cD}$. Finally, in analogy to~\Cref{eq:dfromD}, we introduce the amputated divergence-free amplitude,
    \begin{align}
    \bm{m}_{\rm df; 3} = \bm{\cM}_2^{-1} \, \bm{\cM}_{\rm df,3} \, \bm{\cM}_2^{-1} \, .
    \label{eq:mfromM}
    \end{align}

In our notation, all of the above objects have implicit generalized matrix indices, which are summed or integrated over in products. These indices are exemplified by
    \begin{equation}
    [\bm{\cM}_3]_{ip\ell' s'; j k \ell s} \equiv
    \cM_{3;\ell' s';\ell s}^{(ij)J}(p,k)\,,
    \label{eq:explicit_indices}
    \end{equation}
where the quantity on the right-hand side gives a more explicit expression in which discrete and continuous variables are distinguished. Here and throughout, we keep dependence on the total energy, $E$, implicit, 
and set the total momentum to $\bm P=0$.
The set-up here is that a three-particle system can be described by first choosing one of the particles as a ``spectator'', with the remaining two being the ``pair''. The indices $i,j\in \{D,\pi\}$ denote the spectator flavor.
The momenta of the final and initial spectators are $\bm p$ and $\bm k$, respectively, with magnitudes $p$ and $k$. The choice of spectator momentum determines the invariant mass of the pair. For example, in the final state, the squared invariant mass is
    \begin{equation}
    \sigma_i(p) = [E-\omega_i(p)]^2 - p^2\,,\qquad
    \omega_i(p) = \sqrt{m_i^2 + p^2}\,.
    \label{eq:sigmadef}
    \end{equation}
The dependence on the direction of the members of the pair in its center-of-momentum (c.m.) frame is decomposed into spherical harmonics, with $s'$ and $s$ being, respectively, the ``spin'' of the final and initial pair. The relative angular momentum of the pair and spectator in the overall c.m.~frame is given by $\ell'$ and $\ell$, respectively, for the final and initial state. Finally, the amplitude is projected onto total angular momentum $J$, following the procedure described in refs.~\cite{Jackura:2023qtp,\DRS}. Here, we consider the case $J=1$—the channel of the $T_{cc}^+$—and drop the superscript $J$ on $\bm{\cM}_3$ and other quantities.

Multiplication of generalized matrices is defined by
    \begin{align}
    \left[\,\bm{\cA} \, \bm{\cB} \,\right]_{i p \ell' s'; j k \ell s} &= 
    \sum_{n=D,\pi} \sum_{\ell''_n = 0}^{\ell_{\rm max}^{(n)}} \sum_{ s''_n = 0}^{s_{\rm max}^{(n)}} 
    \int_{q_n} 
    \cA^{(i n)J}_{\ell' s';\ell''_n s''_n}(p,q_n) \, 
    \cB^{(n j)J}_{\ell''_n s''_n;\ell s}(q_n,k) \,,
    \label{eq:3-multip}
    \end{align}
In principle, the sums over intermediate $\ell$ and $s$ run up to infinity. In practice, they must be truncated, as indicated by the upper limits on the respective sums. We discuss this further below.
Integrals over on-shell momenta involve the relativistically-invariant measure,
    \begin{align}
    \int_{q_n} &\equiv \int 
    \frac{dq_n \, q_n^2}{2 \pi^2} \frac1{2 \omega_n(q_n)} \,.
    \label{eq:generalized-matrix}
    \end{align}
In the RFT formalism, such integrals are regularized by a smooth cutoff function contained in the amplitudes such as $\cA$ and $\cB$. The form of this cutoff depends on the intermediate channel $n$, and is determined by the positions of the left-hand singularities in the two-body scattering amplitudes of the corresponding pair~\cite{\DRS}. 
The smooth nature of the cutoff is a requirement in the finite-volume RFT step, but in infinite volume one can use a hard cutoff. This facilitates comparisons between different approaches, and thus, in the main text, we use a hard cutoff throughout. We discuss the relation between smooth and hard cutoffs in \Cref{app:cutoff}.

It is important to emphasize that $\bm{\cM}_3$ is not the complete amplitude, but rather an unsymmetrized version thereof.\footnote{%
In previous work in the RFT literature, it has been denoted $\bm{\cM}_3^{(u,u)}$, but here, following ref.~\cite{\DRS}, we drop the superscript since we only need to consider the unsymmetrized quantity.}
Its definition is most simply given in the language of time-ordered perturbation theory (TOPT), as described in refs.~\cite{\BSQC,\BSnondegen,\BStwoplusone,\tetraquark}. For a given TOPT diagram contributing to the full amplitude, the final- (initial-) state spectators are determined by the latest (earliest) two-particle interaction, which, by definition, always involves the final state (initial state) pair. If one or both of these interactions involve three particles, then the diagram is partitioned in a well-defined way between the different spectator choices. In this manner, the full set of TOPT diagrams is partitioned into the different unsymmetrized amplitudes. The full amplitude can be reconstructed by combining these in a ``symmetrization'' procedure described in detail in ref.~\cite{\tetraquark}.

We now return to the choice of maximal $\ell$ and $s$. In ref.~\cite{\DRS}, we used $s_{\rm max}=1$ and $\ell_{\rm max}=2$ in all channels.
Given that $J=1$, these restrictions lead to three allowed $(\ell,s)$ combinations for $D\pi$ pairs
($D$ spectators) and one for $DD$ pairs ($\pi$ spectators), 
    \begin{equation}
    (\ell_D,s_D) = (0,1),\ (2,1),\ (1,0)\;\qquad  (\ell_\pi,s_\pi) = (1,0)\,.
    \end{equation}
Note that, although the $p$-wave ($s_D=1$) $D^*$ channel for external pairs (with $D$ as the spectator) is the most significant, the $s_D=0$ pair-spectator states contribute in intermediate states, as does the $(\ell_\pi,s_\pi)=(1,0)$ channel (which has $\pi$ as the spectator).


\subsection{Restricted three-body amplitude}
\label{sec:RFT:restricted}

In the interests of simplicity of presentation, we now restrict amplitudes to the $s$-wave $D\pi$ channel with $(\ell_D,s_D)=(0,1)$, and assume that the contributions from other channels are small. This restriction is achieved by setting $\bm{\cM}_2$ and $\bm{\cK}_3$ to zero in the corresponding channels. We note that analyses using the LS equation also have considered this restricted system~\cite{Du:2023hlu, Meng:2023bmz, Collins:2024sfi, Abolnikov:2024key, Vujmilovic:2024snz}. We also observe that this truncation provides a good numerical approximation to the full solution~\cite{\DRS}.

The above-described restriction leads to a single-channel problem. In this section, we write out the corresponding simplified equations in full detail, in preparation for NR reduction.

\subsubsection{Ladder amplitude}

We first consider the ladder amplitude. The integral equation \Cref{eq:inteqd}, restricted to the $(\ell_D,s_D)=(0,1)$ channel, becomes
    \begin{align}
    d(p,k) = 
    - G(p,k) 
    - \int_{q} 
    G(p,q) \, 
    \cM_{2}(q) \, 
    d(q,k) \,.
    \label{eq:simplified_ladder}
    \end{align}
Above, and in the following text, we simplify the notation by dropping the superscripts indicating spectator flavor, which is always $D$, the total angular momentum, always $J=1$, and the subscripts denoting external $(\ell, s)$ angular momenta, always $(0,1)$. 
Thus, for example, $d_{01;01}^{(DD)J=1}(p,k) \to d(p,k)$. In the following, we also drop flavor indices from momenta (e.g., $p_D \to p$), energies (e.g., $\omega_D(p) \to \omega(p)$), and two-body invariant masses (e.g., $\sigma_D(p) \to \sigma(p)$). The projected OPE exchange kernel, $G(p,k) \equiv G_{01;01}^{(11)}(p,k)$, 
is given by eq.~(B.6) of ref.~\cite{\DRS},
    \begin{align}
    \begin{split}
    G(p,k) &= 
    \frac{H(p) H(k)}{2 q^\star(p) q^\star(k)} \Bigg[ \frac{1}{3} 
    \left\{ g_{kp} (\gamma_p + 2) 
    + g_{pk} (\gamma_k + 2) \right\} \, Q_0(z) \\ 
    &\ \ \ + \left\{ g_{pk} g_{kp} + \frac{1}{5} \left( -2 + 2 \gamma_p + 2 \gamma_k + 3 \gamma_p \gamma_k \right) \right\}  \, Q_1(z) \\ 
    &\ \ \ + \frac{2}{3} \left\{ g_{kp} (\gamma_p - 1) + g_{pk} (\gamma_k - 1) \right\} \, Q_2(z) 
    + \frac{2}{5} (\gamma_k - 1)(\gamma_p - 1) \, Q_3(z) \Bigg] \, .
    \end{split}
    \label{eq:G0101}
    \end{align}
Here $Q_i(z)$ are Legendre functions of the second kind, whose argument is given by
    \begin{align}
    z = \frac{ [E - \omega(p) - \omega(k)]^2 - p^2 -k^2 - m^2_{\pi} }{2pk}\,,
    \label{eq:zdef}
    \end{align}
while the relativistic factors are
    \begin{align}
    \beta_p = \frac{p}{E - \omega(p)} \, , ~~~ \gamma_p = \frac{1}{\sqrt{1-\beta_p^2}} \, , 
    ~~~ g_{pk} = \frac{\beta_p \gamma_p \omega_k }{k} \,,
    \label{eq:relativistic-factors}
    \end{align}
along with analogous formulas under $k\leftrightarrow p$ exchange. The quantities $q^\star(p)$ [$q^\star(k)$] are the magnitudes of the momenta of the particles forming the final (initial) $D\pi$ pair, evaluated in the c.m.~frame of the corresponding pair. They are given by
    \begin{align}
    q^\star(p) = \frac{\lambda^{1/2}(\sigma(p), m_D^2, m_\pi^2)}{2\sqrt{\sigma(p)}} \,,
    \label{eq:qqstars}
    \end{align}
where $\lambda(x,y,z)$ is the K\"allen triangle function. $H(p)$ is a smooth cutoff function, which, as noted above, will be replaced here by a hard cutoff at $q=\Lambda$.

The remaining quantity in \Cref{eq:simplified_ladder} is the on-shell $p$-wave $D\pi$ amplitude, $\cM_{2}(q)$.
This can be expressed in terms of the two-particle K matrix,
    \begin{align}
    \cM_{2}(q) = \frac{1}{ \cK_2^{-1}(q) - i \rho(q) } \, .
    \label{eq:MfromK}
    \end{align}
In ref.~\cite{Dawid:2024dgy}, a truncated effective range expansion was used for $\cK_2$,
    \begin{align}
    \cK_2^{-1}(p) = \frac{1}{8 \pi \sqrt{\sigma(p)} [q^\star(p)]^2 } \left( - \frac{1}{a} + \frac{1}{2} \, r^2 \, [q^\star(p)]^2 \right) \, , \quad \rho(p) = \frac{q^\star(p)}{8 \pi \sqrt{\sigma(p)}} \,,
    \label{eq:2b-kmatrix}
    \end{align}
with the scattering length, $a$, and the effective range, $r$, chosen such that the amplitude exhibited the below-threshold pole corresponding to the $D^*$ bound state. Close to this pole, the amplitude takes the form
    \begin{align}
    \cM_{2}(q) \simeq \cM_{2}^p(q) = \frac{\zeta^2}{2 m_{D^*}}
    \frac{1}{ \sqrt{\sigma(q)} - m_{D^*} + i\varepsilon } \,,
    \label{eq:M2pole}
    \end{align}
where the residue is determined by~\cite{\DRS}
    \begin{align}
    \zeta^2 = - \frac{1}2 \, q_0^2 \, g_{DD^*\pi}^2 = - q_0^2 \, g^2 \,  \frac{2m_Dm_{D^*}}{f_\pi^2} \, .
    \label{eq:zeta}
    \end{align}
Here $g_{DD^*\pi}$ is the standard $DD^*\pi$ coupling constant~\cite{Belyaev:1994zk}, while $g$ is the coupling used in heavy-quark effective theory, with $f_\pi\approx 130\;$MeV. The quantity $q_0$ is the momentum of the $D$ and $\pi$ in the $D^*$ c.m.~frame at the $D^*$ pole,
    \begin{align}
    q_0 = \frac{\lambda^{1/2}(m_{D^*}^2, m_D^2, m_\pi^2)}{2 m_{D^*}} \,.
    \label{eq:q0def}
    \end{align}
Note that $q_0$ is purely imaginary, implying that $\zeta^2$ is real and positive.

\subsubsection{Divergence-free amplitude}

We now include the impact of including a nonzero $\cK_3$ within the system in which only the $p$-wave $D\pi$ subchannel contributes. Thus we consider only the component $\cM_{{\rm df},3}(p,k) \equiv \cM^{(DD)J=1}_{{\rm df},3;01;01}(p,k)$ of the divergence-free amplitude, and drop superscripts and subscripts henceforth. From \Cref{eq:LTR}, we have
    \begin{align}
    \cM_{{\rm df}, 3}(p,k) = 
    \int_{q}\int_{r}
    \cL(p,q) \,
    \cT(q,r) \, 
    \cR(r,k) \,.
    \label{eq:Mdf_reduced}
    \end{align}
The endcaps are
    \begin{align}
    \cL(p,q) &= 
    \left[\tfrac13 - \cM_{2}(p) \, \widetilde\rho(p) \right] 
    \bar\delta(p-q)
    - \cD(p,q) \, 
    \widetilde\rho(q)\,,
    \label{eq:cL0101}
    \\
    \cR(p,q) &= 
    \bar\delta(p-q) \left[\tfrac13 - 
    \widetilde\rho(q) \, \cM_{2}(q) \right] 
    - \widetilde\rho(p) \, \cD(p,q) \,,    
    \label{eq:cR0101}
    \end{align}
where
    \begin{align}
     \bar\delta(p-q) = (2\pi)^2 \, \frac{\omega_p}{p^2} \, \delta(p-q) \,,
     \label{eq:deltafcn}
    \end{align}
and the modified phase space factors are given by
    \begin{align}
    \widetilde\rho(p) = H(p) \, 
    \frac{q^\star(p)}{8\pi \sqrt{\sigma(p)}} 
    \left(-i + \frac{c_{\rm PV}}{[q^{\star}(p)]^3} \right) \, .
    \label{eq:mod-rho}
    \end{align}
Here $c_{\rm PV}$ is a (dimensionful) constant that must be introduced in the RFT finite-volume quantization condition for the derivation to remain valid in the presence of a bound-state pole~\cite{Romero-Lopez:2019qrt}. The result for $ \cK_3$ that one extracts from a finite-volume analysis will depend on $c_{\rm PV}$, and, thus, one must also keep the $c_{\rm PV}$ term when solving the integral equation to obtain the amplitude $\cM_3$.

Following ref.~\cite{\tetraquark}, we describe $\bm{\cK}_3$ using an expansion about the $DD\pi$ threshold, working to linear order. While there are four terms at this order, only one contributes to the $I=0$, $J=1$ channel, and it is given by~\cite{\DRS},
    \begin{align}
    \cK_{3}(p,k) &= 
    \frac{2}{27} \, \cK_3^E \, \frac{1}{M^2} q^\star(p) \, (\gamma_p + 2) 
    q^\star(k) ( \gamma_k + 2) \, ,
    \label{eq:K3def}
    \end{align}
where $M=2 m_D + m_\pi$, and $\cK_3^E$ is an {\em a priori} unknown coefficient. This form is manifestly separable,
    \begin{equation}
    \cK_{3}(p,k) = \cK_L(p) \, \cK_R(k)\,,
    \quad \cK_L(p) = \cK_R(p) = \sqrt{\frac{2\cK_3^E}{27}} \frac{q^\star(p)}{M} (\gamma_p+2) \,,
    \end{equation}
and so the integral equation that determines $\bm{\cT}$, \Cref{eq:T}, can be solved straightforwardly~\cite{\DRS},
    \begin{align}
    \cT(p,k) &= 
    \cK_L(p) \,
    \frac{1}{1 + \cI} \, 
    \cK_L(k) \, ,
    \label{eq:Tres}
    \end{align}
where the integral is given by
    \begin{align}
    \begin{split}
    \cI &= \int_{q} \cK_L(q) \,
    \widetilde\rho(q) 
    \left[\tfrac13 - \cM_{2}(q) \, 
    \widetilde\rho(q) \right] 
    \cK_L(q)
    \\
    &\ \ - \int_{q} \int_{r} 
    \cK_L(q) \,
    \widetilde\rho(q) \, 
    \cM_{2}(q) \,
    d(q,r) \,
    \cM_{2}(r) \,
    \widetilde\rho(r) \,
    \cK_L(r)\,,
    \end{split}
    \label{eq:cIfull}
    \end{align}

\subsection{LSZ reduction}
\label{sec:LSZ}

To complete the above discussion, we recall how the LSZ reduction formula is used to obtain the $s$-wave $DD^*$ scattering amplitude~\cite{Jackura:2020bsk, Dawid:2023kxu},
    \begin{align}
    \cM_{DD^*}^{\ell=0}(E) & =
    \lim_{\sigma(p), \, \sigma(k) \to m_{D^*}^2 } - \frac{1}{\zeta^2}
    \left( {\sigma(p) -m_{D^*}^2}\right) \, 
    \cM_{3}(p,k) \, 
    \left( {\sigma(k) -m_{D^*}^2}  \right)
    \label{eq:LSZ} \\
    & = -\zeta^2 \, \big[ d(p_0,p_0)
    + m_{\rm df,3}(p_0,p_0) \big] \, ,
    \label{eq:LSZ-2}
    \end{align}
where, in the second line, we have used the fact that  $d$ and $m_{\rm df,3}$ are amputated,
\Cref{eq:dfromD,eq:mfromM}. The quantity $p_0$ is the magnitude of the spectator momentum that, for a given energy, sets the pair invariant mass-squared $\sigma(p)$ equal to $m_{D^*}^2$,
    \begin{equation}
    p_0 = \frac{\lambda^{1/2}(E^2, m_{D^*}^2, m_D^2)}{2E}\,.
    \label{eq:p0def}
    \end{equation}
The zeros in $[\cM_{2}^{(1)}]^{-1}$ remove the $1/3$ terms in $\cL$ and $\cR$, \Cref{eq:cL0101,eq:cR0101}, leading to the simplified form for the amputated divergence-free contribution when external momenta are set to $p_0$,
    \begin{align}
    m_{\rm df,3}(p_0,p_0) &= 
    L(p_0) \, \frac1{1+\cI} \, L(p_0)\,,
    \label{eq:simplified-mdf3}
    \\
    L(p_0) &= \widetilde 
    \rho(p_0) \, 
    \cK_L(p_0)
    + \int_{q} d(p_0,q) \,
    \cM_{2}(q) \, \widetilde\rho(q) \, 
    \cK_L(q) \, . 
    \label{eq:Lp0} 
    \end{align}
The equality between left and right endcaps in this expression relies on the symmetry of $d(p,q)$, which in turn follows from that of $G(p,k)$.

\section{Lippmann-Schwinger approach with Chiral EFT potential}
\label{sec:LSE}

As noted in the introduction, several studies of the $T_{cc}^+$ use the LS equation to describe the $DD^*$ system, with potentials determined by chiral EFT and three-body effects included in the sense of ref.~\cite{Baru:2011rs}. Specifically, the effects of one-pion exchanges are incorporated in the equation by deriving it from a more general description of the coupled $DD^*$ and $DD\pi$ system. This allows the inclusion of three-body intermediate states in the integral equation for the elastic $DD^*$ amplitude. $DD\pi$ propagation contributes to the non-zero $D^*$ width in the resulting generalized LS equation. In the case of a heavy pion and stable $D^*$, the formalism takes the form of the standard LS equation, with the potentials describing virtual pion exchange and a set of contact interactions. We note that this approach does not include three-body potentials or pairwise $D\pi$ and $D D$ potentials.\footnote{%
A general description of $DD\pi$ scattering in the form of an off-shell integral equation could include these as described, e.g., in ref.~\cite{Stadler:1991zz}. }
In this respect, it differs from the (unrestricted) RFT approach in which these two types of interaction are incorporated through the three-body K matrix, $\bm \cK_3$, and the general form of the two-particle amplitude, $\bm \cM_2$, respectively.

We will explicitly compare the RFT formalism with the formulas presented in ref.~\cite{Collins:2024sfi}, where only the $s$-wave $DD^*$ channel is included. This matches the approximations made to arrive at the restricted three-body amplitude described in \Cref{sec:RFT:restricted}. The LS equation is [eq.~(21) of ref.~\cite{Collins:2024sfi}, with notation slightly changed]
    \begin{align}
    T(\bm p',\bm p;E) &= V(\bm p', \bm p; E) +
    \int_0^\Lambda \frac{d^3q}{(2 \pi)^3} \, V(\bm p', \bm q; E) 
    \, \cG(q;E) \, T(\bm q, \bm p;E) \, ,
    \label{eq:LS}
    \end{align}
where the resolvent is [this has the opposite sign to $G(\bm q; E)$ of eq.~(22) of ref.~\cite{Collins:2024sfi}, and has the sign of $i\epsilon$ corrected]
    \begin{align}
    \cG(q;E) = \frac{1}{E_{\rm NR} - \frac{q^2}{2 \mu} + i \epsilon} \,,
    \qquad
    E_{\rm NR} = E - m_D - m_{D^*}\,,
    \label{eq:Gbar}
    \end{align}
with $\mu$ the reduced mass of the $D^* D$ system, and $\Lambda$ the cutoff. We work in the c.m.~frame and use the notation that $\bm p'$ and $\bm p$ are, respectively, the final and initial momenta of the $D$ mesons. The scattering matrix $T$ is in general off shell; to obtain the physical amplitude, one must consider elastic scattering, in which $p'=|\bm p'|$ and $p=|\bm p|$ are equal and given by $p_0^{\rm NR}$, which is related to the energy by
    \begin{equation}
    E_{\rm NR} = \frac{(p_0^{\rm NR})^2}{2\mu}\,.
    \label{eq:on-shell-mom}
    \end{equation}
The use of the nonrelativistic expressions for the energies of the $D^*$ and $D$ is standard in applications of the LS equation.

The potential used in ref.~\cite{Collins:2024sfi} consists of a one-pion exchange contribution $V_\pi$ and a contact contribution $V_{\rm CT}$,
    \begin{align}
    V(\bm p',\bm p; E) =  V_\pi(\bm p', \bm p; E) + V_{\rm CT}(\bm p',\bm p) \, .
    \label{eq:Vdecomp}
    \end{align}
The former is given by
    \begin{align}
    V_\pi(\bm p',\bm p;E) &= 
    3 \frac{g^2}{2 f_\pi^2} \, 
    (\bm\epsilon'^*\cdot \bm q) \, 
    (\bm q \cdot \bm\epsilon)
    \left[\frac1{D_{DD\pi}}+\frac1{D_{D^*D^*\pi}} \right] \, ,
    \label{eq:Vpi}
    \end{align}
which was reconstructed from eqs.~(2) and (18) of ref.~\cite{Collins:2024sfi}, taking into account the difference in normalization of $f_\pi$ used in that work. Here, $\bm q= \bm p'+\bm p$, $\bm \epsilon'$ ($\bm \epsilon$) are the final (initial) $D^*$ polarizations, and the energy denominators (given the same names as in ref.~\cite{Collins:2024sfi}) arise from TOPT,
    \begin{align}
    D_{DD\pi} &= {2\omega_\pi{(q)}} \left({E_{\rm NR} + \Delta - \frac{p'^2 + p^2}{2 m_D} - \omega_\pi(q)}\right)\,,
    \label{eq:DDDpi}
    \\
    D_{D^*D^*\pi} &= {2\omega_\pi{(q)}} \left({E_{\rm NR} - \Delta - \frac{p'^2+p^2}{2 m_{D^*}} - \omega_\pi{(q)}}\right)\,,
    \label{eq:DDstDstpi}
    \end{align}
where 
    \begin{equation}
    \Delta = m_{D^*}- m_D\,.
    \label{eq:Deltadef}
    \end{equation}
The contact contribution to the potential is
    \begin{align}
    V_{\rm CT}(\bm p',\bm p) &= 2 c_0 + 2 c_2 (p'^2+p^2) \, ,
    \label{eq:VCT}
    \end{align}
where $c_0,c_2$ are constants parameterizing unknown short-range interactions. Note that a more general potential has been considered in  recent work~\cite{Prelovsek:2025vbr}.

\subsection{One-pion exchange contribution}
\label{sec:LS-OPE}

In this section, we study the solution to the LS equation, \Cref{eq:LS}, setting $V=V_\pi$, which is analogous to considering only the ladder contribution $\bm{\cD}$ in the RFT formalism. We denote the resulting solution $T_\pi$.

We first project the LS equation onto the $J=1, L=0$ channel. Denoting this channel with the subscript $S$, one finds
    \begin{align}
    T_{\pi,S}(p',p;E) = V_{\pi,S}(p',p;E) + \int_0^\Lambda \
    \frac{dq \, q^2}{2 \pi^2} \, V_{\pi,S}(p',q;E) \, \cG(q;E) \, T_{\pi,S}(q,p;E) \, ,
    \label{eq:LSproj}
    \end{align}
Here
    \begin{align}
    V_{\pi,S}(p',p;E) &= \frac13 \int \frac{d\Omega_{p'}}{4\pi} \frac{d\Omega_{p}}{4\pi} \, \sum_{\epsilon',\epsilon} \, 
    (\bm \epsilon'\cdot \bm \epsilon^*) \,
    V_{\pi}(\bm p',\bm p;E) \,,
    \end{align}
where the sum runs over the final and initial polarizations, and $(\bm \epsilon'\cdot \bm \epsilon^*)$ is the projector onto the $J=1, L=0$ channel~\cite{Baru:2019xnh}.
Using the NR result
    \begin{equation}
    \sum_\epsilon \epsilon^*_i \epsilon_j = \delta_{ij}\,,
    \end{equation}
one obtains
    \begin{align}
    V_{\pi,S}(p',p;E) &= \frac{g^2}{2 f_\pi^2}
    \int \frac{d\Omega_{p'}}{4\pi} \frac{d\Omega_{p}}{4\pi} \, q^2 
    \left[\frac1{D_{DD\pi}}+\frac1{D_{D^*D^*\pi}} \right]
    \label{eq:VpiS}
    \end{align}

To proceed, we introduce the NR power counting that we will use to compare the LS and RFT formalisms. 
To motivate our choice, we note that, in the lattice calculation of ref.~\cite{Padmanath:2022cvl}, the $D^*-D$ mass splitting satisfies $\Delta \approx m_\pi/2$,  with $m_\pi/M_D\approx 0.15$. Furthermore, the energy range of interest (see \Cref{fig:phase-shifts} below) has $E_{\rm NR}/m_D$ between $-0.01$ and  $0.04$. We thus choose our power counting to be
    \begin{equation}
    \frac{\Delta}{m_D} \sim \frac{m_\pi}{m_D} \sim \epsilon\,,\quad
    \frac{E_{\rm NR}}{m_D} \sim \frac{p^2}{m_D^2} \sim \frac{p^2}{m_{D^*}^2} \sim \epsilon^2 \,. 
    \label{eq:power-counting}
    \end{equation}
Keeping the leading term in this NR expansion in the denominators \Cref{eq:DDDpi,eq:DDstDstpi} yields
    \begin{align}
    D_{DD\pi} &\approx 2 \omega_\pi(q) [\Delta - \omega_\pi(q)]
    = \frac{2 \omega_\pi(q)}{\Delta + \omega_\pi(q)}
    (\Delta^2 -m_\pi^2 - q^2)\,,
    \label{eq:DDDpNR}
    \\
    D_{D^*D^*\pi} &\approx 2 \omega_\pi(q)[-\Delta - \omega_\pi(q)]
    = -\frac{2 \omega_\pi(q)}{\Delta - \omega_\pi(q)}
    (\Delta^2 -m_\pi^2 - q^2)\,,
    \label{eq:DDstDstpNR}
    \end{align}
which combine to give
    \begin{align}
    \left[\frac1{D_{DD\pi}}+\frac1{D_{D^*D^*\pi}} \right] &\approx
    - \frac1{q^2+m_\pi^2 - \Delta^2}\,.
    \end{align}
We stress that this quantity does not diverge since,
given the stability of the $D^*$ meson, we must have $\Delta < m_\pi$.

We can now perform the integral in \Cref{eq:VpiS}, obtaining
    \begin{align}
    V_{\pi,S}(p',p;E) &\approx - \frac{g^2}{2 f_\pi^2}
    \int_{-1}^{1} \frac{dc_\theta}2
    \frac{p^{\prime\,2}+p^2 + 2 p' p c_\theta}{p^{\prime\,2}+p^2 + m_\pi^2-\Delta^2 + 
    2 p'p c_\theta} \equiv V_{\pi,S}^{\rm NR}(p',p) \, ,
    \label{eq:VpiSb}
    \\
    V_{\pi,S}^{\rm NR}(p',p) &=  \frac{g^2}{2 f_\pi^2} \left[ \frac{p^{\prime\,2}+ p^2}{2p'p} Q_0(z_{\rm NR}) + Q_1(z_{\rm NR}) \right] \,,
    \label{eq:VpiSc}
    \end{align}
where $z_{\rm NR}$ is
    \begin{align}
    z_{\rm NR} &=\frac{\Delta^2 - p'^2 -p^2 -m_\pi^2}{2 p' p} \,.
    \label{eq:zNR}
    \end{align}

To make the comparison with the RFT results, which will be done in the following section, we must take care of the relative normalization of the NR scattering amplitude $T_S(p',p;E)$ and the relativistic amplitude $\cM_{DD^*}$. To do so, we note that the $s$-wave amplitude can be expressed in terms of the phase shift by
(see eq.~(12) of ref.~\cite{Collins:2024sfi})
    \begin{align}
    T^{\ell=0}(q^\star) 
    &= - \frac{2\pi}{\mu} \frac1{q^\star \cot \delta(q^\star) - iq^\star}\,,
    \end{align}
where $q^\star$ is the momentum of both $D$ and $D^*$ in the overall c.m. frame.\footnote{%
In the relativistic case, this is given by $p_0$, \Cref{eq:p0def}.}
This should be compared to the relativistic result
    \begin{align}
    \cM_{DD^*}^{\ell=0}(q^\star) &= 
    \frac{8\pi E_{\rm on}}{q^\star \cot \delta(q^\star) - i q^\star}\,,\quad
    E_{\rm on} = \sqrt{m_{D^*}^2+q^{\star 2}} + \sqrt{m_{D}^2+q^{\star 2}}\,,
    \end{align}
from which we learn that the relative normalization is
    \begin{equation}
    \cM_{DD^*}^{\ell=0} = - f_2 \, T \,,
    \qquad f_2 = (4\mu E_{\rm on}) \,.
    \label{eq:MvsTNR}
    \end{equation}
Thus, in the following, we introduce the rescaled LS amplitudes
    \begin{equation}
    \overline{T}_{\pi,S}(p',p;E) \equiv - f_2 T_{\pi,S}(p',p;E)\ \ {\rm and}\ \
    \overline{T}_{S}(p',p;E) \equiv - f_2 T_{S}(p',p;E)\,,
    \label{eq:renormT}
    \end{equation}
which will be compared directly to the corresponding quantities in the RFT results. In particular, the result \Cref{eq:LSproj} becomes
    \begin{align}
    \overline{T}_{\pi,S}(p',p;E) \approx -f_2 V_{\pi,S}^{\rm NR}(p',p) + \int_0^\Lambda \
    \frac{dq \, q^2}{2 \pi^2} \, V_{\pi,S}^{\rm NR}(p',q) \, \cG(q;E) \, \overline{T}_{\pi,S}(q,p;E) \, ,
    \label{eq:LSprojNR}
    \end{align}
after the NR approximations have been made.

\subsection{Full amplitude }
\label{sec:LS-full}

We now consider the solution to the LS equation keeping the full potential, i.e., both $V_\pi$ and $V_{\rm CT}$. It is shown in appendix B of ref.~\cite{Collins:2024sfi} that the solution is
    \begin{align}
    T(\bm p', \bm p;E) &= T_\pi(\bm p', \bm p;E) + \sum_{a,b=1}^{2} 
    X_a(\bm p';E) \, 
    \hat t_{ab} \, X_b(\bm p;E) \,,
    \\
    X_a(\bm p;E) &= f_a(p) + \int_0^\Lambda \frac{d^3q}{(2\pi)^3} \, 
    T_\pi(\bm p, \bm q;E) \, 
    \cG(q;E) \, 
    f_b(q) \, ,
    \\
    [\hat t^{-1}]_{ab} &= [\hat C^{-1}]_{ab} + \overline{\cI}_{ab} \,,
    \\
    \overline{\cI}_{ab} &= - \int_0^\Lambda \frac{d^3q}{(2\pi)^3} f_a(q) \,
    \cG(q;E) \, 
    X_b(\bm q;E) \ , 
    \end{align}
where $a,b$ are matrix indices running over $1$ and $2$, with repeated indices summed as usual, and the vector $f$ and matrix $\hat C$ are
    \begin{equation}
    f(p) = \begin{pmatrix} 1 \\ p^2 \end{pmatrix}\,,\quad
    \hat C = 2 \begin{pmatrix} c_0 & c_2 \\ c_2 & 0 \end{pmatrix}
    \,,
    \end{equation}
with the constants $c_0$ and $c_2$ appearing in $V_{\rm CT}$, \Cref{eq:VCT}. Some signs in the results above differ from those in ref.~\cite{Collins:2024sfi} because we define the resolvent $\cG$ with the opposite sign.

To proceed, we drop the $c_2$ term, which is suppressed in the NR limit, project onto the $J=1, L=0$ channel, evaluate the amplitude on shell, and convert to relativistic normalization using \Cref{eq:renormT}. This leads to the final form of the LS equation,
    \begin{align}
    \overline{T}_S(p_0^{\rm NR}, p_0^{\rm NR};E) &= \overline{T}_{\pi,S}(p_0^{\rm NR}, p_0^{\rm NR};E) - 4m_D^2\, X_S(p_0^{\rm NR};E)\ t\  X_S(p_0^{\rm NR};E)\,,
    \label{eq:TSbarfinal}
    \end{align}
    where $p_0^{\rm NR}$ is the on-shell momentum given in \Cref{eq:on-shell-mom}, and
    \begin{align}
    X_S(p_0^{\rm NR};E) &= 1 - 
    \frac1{4m_D^2} \int_0^\Lambda \frac{dq\, q^2 }{2\pi^2} \, 
    \overline{T}_{\pi,S}(p_0^{\rm NR}, q;E) \,
    \cG(q;E) \,,
    \label{eq:XSfinal}
    \\
    t &= \frac{2 c_0}{1+ 2 c_0 \, \overline{\cI}_{\rm NR}} \, ,
    \label{eq:tfinal}
    \\
    \overline{\cI}_{\rm NR} &= - \int_0^\Lambda \frac{dq \,q^2}{2\pi^2} \,
    \cG(q;E) \left[ 1 -
    \frac1{4m_D^2}
    \int_0^\Lambda \frac{dr\, r^2}{2\pi^2} \,
    \overline{T}_{\pi,S}(q, r;E) \,
    \cG(r;E)
    \right] \,.
    \label{eq:IbarNRfinal}
    \end{align}

\section{Comparison between RFT and LS approaches}
\label{sec:NR-RFT}

In this section, we describe and apply the two approximations that are needed to bring the RFT equations into agreement with the LS forms.
The first approximation is to take the NR limit of quantities entering the equations; the second is assuming dominance of the pole contribution to the $D\pi$ amplitude $\cM_2$.

Our NR approximation follows and systematizes the approach of refs.~\cite{Jackura:2019bmu, Dawid:2020uhn}.
We use the power counting of \Cref{eq:power-counting}, extended to include $q_0$ [from \Cref{eq:q0def}] and $p_0$,
    \begin{equation}
    \frac{\Delta}{m_D} \sim \frac{m_\pi}{m_D} \sim \epsilon\,,\quad
    \frac{|q_0^2|}{m_D^2} \sim \frac{m_\pi^2}{m_D^2} \sim \frac{p_0^2}{m_D^2} \sim  \frac{E_{\rm NR}}{m_D} \sim \epsilon^2 \, , 
    \label{eq:power-counting2}
    \end{equation}
We note that \Cref{eq:q0def,eq:p0def} imply
    \begin{align}
    q_0^2 &= (\Delta^2-m_\pi^2) \left[1 + \cO(\epsilon)\right]
    \ \ {\rm and}\ \
    p_0^2 = E_{\rm NR} m_D \left[1 + \cO(\epsilon) \right]
    = (p_0^{\rm NR})^2 \left[1 + \cO(\epsilon) \right]\,,
    \label{eq:NRmomenta}
    \end{align}
which shows the consistency of the power counting.

We carry out the reduction of the three-body equations by replacing the amplitudes---such as $\cM_2$ and $G$---with their NR approximations within the three-body integral equations. While this procedure involves a consistent NR treatment at the level of integrands, it does not correspond to taking the NR limit of the full relativistic solution. The difference between these two procedures arises from momenta close to the cutoff scale $\Lambda$,
for which the power-counting scheme of \Cref{eq:power-counting2}
can break down. 

In particular, we use $p_{\rm max}=\Lambda=500\;$MeV in our numerical examples below, for which $(p_{\rm max}/m_D)^2 = 0.067$ which is close to $\Delta/m_D=0.063$.
Thus at the highest momenta we have $p^2/m_D^2 \sim \epsilon$ rather than $\epsilon^2$ as assumed in \Cref{eq:power-counting2}. However, conceptually, the error associated with ignoring this change of power counting is the same as that made in the LS framework, for the latter employs NR potentials and propagators inside the integral, even for high cutoff scales. Therefore, although we do not quantify relativistic corrections rigorously, our reduction strategy is appropriate for relating the structure of the three-body formalism to the LS equation.

Before proceeding to the details, we recall here, for the sake of clarity, the
 RFT result for the $DD^*$ amplitude, \Cref{eq:LSZ-2}, 
    \begin{align}
    \cM_{DD^*}^{\ell=0}(E) 
    & = -\zeta^2 \, \big[ d(p_0,p_0)
    + m_{\rm df,3}(p_0,p_0) \big] \,.
    \label{eq:LSZ-3}
    \end{align}
Here $\zeta^2$ is a constant defined in \Cref{eq:zeta}, and the momentum $p_0$ is defined in \Cref{eq:p0def}.
The ladder contribution $d$ satisfies the integral equation \Cref{eq:simplified_ladder}, while $m_{\rm df,3}$ is given by \Cref{eq:simplified-mdf3}.

\subsection{Approximating the RFT ladder amplitude}
\label{sec:RFT-NR-ladder}

We start by considering the RFT ladder amplitude, \Cref{eq:simplified_ladder},
whose building blocks are $G(p,k)$ and $\cM_2(q)$.
For the former, we expand \Cref{eq:G0101} in powers of $\epsilon$ and keep the leading contribution, yielding
    \begin{align}
    \begin{split}
    G(p,k)&= \frac{H(p)H(k)}{q^\star(p) q^\star(k)}
    \left[\frac{k^2+p^2}{2kp} Q_0(z) + Q_1(z) + \cO(\epsilon) \right]
    \\
    &\approx \frac1{q_0^2} 
    \left[\frac{k^2+p^2}{2kp} Q_0(z_{\rm NR}) +  Q_1(z_{\rm NR}) \right]
    \equiv G_{\rm NR}(p,k)\,.
    \end{split}
    \label{eq:GNR}
    \end{align}
Here $z_{\rm NR}$ is the same quantity as appearing in the LS analysis, \Cref{eq:zNR}, with $p'$ replaced here by $k$. To obtain the second line of \Cref{eq:GNR}, we have replaced the smooth cutoff with a hard cutoff, i.e., we set $H(p)=1$ for $p\le\Lambda$, and $H(p)=0$ above this value. Furthermore, we have made the NR approximation $q^\star(p) q^\star(k) \approx q_0^2$, with $q_0$ given in Eq.~\eqref{eq:NRmomenta}.

Comparing \Cref{eq:GNR} to \Cref{eq:VpiSc}, we find that
    \begin{align}
    G_{\rm NR}(p,k) = \frac1{f_1} V_{\pi,S}^{\rm NR}(p,k) \,,
    \qquad
    f_1 =  \frac{q_0^2 g^2}{2 f_\pi^2}\,,
    \end{align}
showing that the driving terms in the integral equations for the ladder amplitude in the RFT and LS approaches are proportional in the NR limit.

We now turn to the $D\pi$ amplitude, $\cM_2(q)$. Here we find that, to match RFT and LS forms, we need to make an additional ``pole dominance'' approximation. Specifically, we must keep only the pole term $\cM_{2}^p$, \Cref{eq:M2pole}, and drop nonpole contributions. This can be justified systematically by expanding the amplitude in a Laurent series around the pole, in which $\cM_2^p$ is the leading order term. The subsequent terms of the expansion provide higher-order contributions; however, in practice, whether neglecting these terms is a reasonable approximation is often a numerical question. We return to this point in \Cref{sec:numerical} and~\Cref{app:offshell}. It is possible to study formally the effect of dropping nonpole terms for the comparison of the three-body RFT with the LS framework. In \Cref{app:offshell}, we describe how the effect of this approximation can be accounted for, in a certain limit, through redefinitions of coupling constants appearing in the resulting LS equations.

Making the pole-dominance approximation, the next step is to take the NR limit of $\cM_{2}^p$, \Cref{eq:M2pole}, leading to
    \begin{align}
    \cM_{2}^p(q) &\approx \frac{\zeta^2}{2 m_{D}} 
    \frac1{E_{\rm NR} - \tfrac{q^2}{2\mu}+ i \epsilon}
    = \frac{\zeta^2}{2m_D} \, \cG(q;E) \, ,
    \label{eq:M2NR}
    \end{align}
where $\cG$ is the NR resolvent introduced in \Cref{eq:Gbar},
and the constant $\zeta^2$ is given in \Cref{eq:zeta}.
Thus the NR resolvent appears in the RFT approach as the NR limit of the pole part of the $D\pi$ amplitude.

Combining the approximate forms in \Cref{eq:GNR,eq:M2NR}, defining 
    \begin{equation}
    d_{\rm NR}(p,k) = - \zeta^2 d(p,k)\,,
    \label{eq:dNRdef}
    \end{equation}
and dropping additional subleading terms in the NR limit arising from the integration measure, we obtain the integral equation
    \begin{align}
    d_{\rm NR}(p,k) \approx
    \zeta^2 G_{\rm NR}(p,k) 
    - \frac1{4m_D^2} \int_0^{\Lambda} 
    \frac{dq \, q^2}{2 \pi^2} \,
    \zeta^2 G_{\rm NR}(p,q) \, 
    \cG(q;E) \,
    d_{\rm NR}(q,k) \,.
    \label{eq:NRladder}
    \end{align}
Combining this with the results
    \begin{equation}
    \zeta^2 = - f_1 f_2 [1 + \cO(\epsilon)] \ \ {\rm and}\ \ 
    f_2 = 4 m_D^2 [1 + \cO(\epsilon)]\,,
    \end{equation}
which follow from the form of $f_2$, \Cref{eq:MvsTNR}, 
we find that the integral equation satisfied by LS amplitude $\overline{T}$ in the NR limit, \Cref{eq:LSprojNR}, is identical to that satisfied by $d_{\rm NR}$ in the same limit,
\Cref{eq:NRladder}, provided we make the identification
    \begin{align}
    d_{\rm NR}(p,k) = \overline{T}_{\pi,S}(p,k;E) \,.
    \label{eq:TpiSvsdNR}
    \end{align}

\subsection{Approximating $m_{\rm df,3}$}
\label{sec:mdf3reduce}

The divergence-free amplitude, \Cref{eq:simplified-mdf3}, contains the integral $\cI$ of \Cref{eq:cIfull}, and the endcap $L(p_0)$, \Cref{eq:Lp0}. The additional NR limits that we need are
    \begin{align}
    \label{eq:KLNR}
    \cK_L(p) &= \cK_R(p) = \sqrt{\frac{\cK_3^E}6} \frac{q_0}{m_D} 
    \left[1+\cO(\epsilon)\right]\,,
    \\
    \widetilde\rho_1^{(1)}(p) &= \frac{-i q_0}{8\pi m_D} 
    \left(1 - \frac{c_{\rm PV}}{|q_0^3|} \right) \left[1+\cO(\epsilon)\right]\,,
    \label{eq:roNR}
    \end{align}
where, in the latter result, we have once again replaced the smooth cutoff with a  hard cutoff. Implementing the pole-dominance approximation in the integrals,
we find the contribution to the $DD^*$ amplitude arising from $\cK_3$ to be,
in the NR limit,
    \begin{align}
    -\zeta^2  m_{\rm df,3}(p_0^{\rm NR},p_0^{\rm NR}) &\approx
    L_{\rm NR}(p_0^{\rm NR}) \, \frac{(\zeta \, c_L)^2}{1+\cI_{\rm NR}} \, L_{\rm NR}(p_0^{\rm NR}) \, .
    \label{eq:mdfNR}
    \end{align}
where
    \begin{align}
    c_L &= \sqrt{\frac{\cK_3^E}6} \frac{(-q_0^2)}{8\pi m_D^2} 
    \left(1 - \frac{c_{\rm PV}}{|q_0^{3}|} \right) \,,
    \label{eq:cL}
    \\
    L_{\rm NR}(p_0^{\rm NR}) &=
    1 - \frac{ 1 }{4 m_D^2} \int_0^\Lambda \frac{dq\, q^2}{2\pi^2} \,
    d_{\rm NR}(p_0^{\rm NR},q) \, 
    \cG(q;E) \, ,
    \label{eq:LNR}
    \end{align}
and
    \begin{align}
    \begin{split}
    \cI_{\rm NR} &= \cI_{\rm NR,0}
    + 
    \frac{(c_L\zeta)^2}{4m_D^2} 
    \int_0^{\Lambda} 
    \frac{dq\, q^2}{2\pi^2} \,
    \cG(q;E) 
    \\
    &\quad -
    \frac{(c_L \zeta)^2}{16m_D^4}
    \int_0^{\Lambda} \frac{dq\, q^2}{2\pi^2}
    \int_0^{\Lambda} \frac{dr\, r^2}{2\pi^2}
    \cG(q;E) \, 
    d_{\rm NR}(q,r)\, 
    \cG(r;E) \, , 
    \end{split}
    \label{eq:cINR}
    \end{align}
with
    \begin{align}
    \cI_{\rm NR,0} &= - \frac{\cK_3^E \Lambda^3}{1728 \, \pi^3 m_D^4 }   
    \big(|q_0^3| - c_{\rm PV} \big) \, .
    \label{eq:cINR0}
    \end{align}
We note that $\cI_{\rm NR,0}$ is an energy-independent constant at this order in the NR expansion.

We are now ready to compare the approximate RFT expression \Cref{eq:mdfNR},
to the corresponding LS expression, which, from \Cref{eq:TSbarfinal,eq:tfinal}, is
    \begin{equation}
    -4 m_D^2 X_S(p_0^{\rm NR};E_0)\frac{2 c_0}{1+ 2 c_0 \, \overline{\cI}_{\rm NR}}  X_S(p_0^{\rm NR};E_0)\,.
    \end{equation}
We first observe, from \Cref{eq:XSfinal,eq:LNR}, and using the identification \Cref{eq:TpiSvsdNR}, that,
    \begin{align}
    X_S(p_0^{\rm NR};E_0) &= L_{\rm NR}(p_0^{\rm NR})\,.
    \end{align}
Furthermore, we have
    \begin{align}
    \overline{\cI}_{\rm NR} &= - 
    \frac{4m_D^2}{c_L^2 \zeta^2} (\cI_{\rm NR} - \cI_{\rm NR,0})\,.
    \end{align}
Using these results, we find that the NR-reduced LS and RFT expressions for the divergence-free contribution agree as long as the following relation 
between constants holds,
    \begin{align}
    c_0 \, m_D^2 = - \frac{1}{8} \frac{c_L^2 \zeta^2}{(1+\cI_{\rm NR,0})}\,.
    \label{eq:matching-c0}
    \end{align}
Since both $c_L^2 $ and $\cI_{\rm NR,0}$ are proportional to $\cK_3^E$, the relation between $c_0$ and $\cK_3^E$ is nonlinear. 

We emphasize that, although our comparison has been limited to LO contact terms, the inclusion of higher-order terms---such as those associated with the $c_2$ coupling in Eq.~(3.9)---is, in principle, straightforward within both frameworks. However, in practice, their inclusion will certainly increase the complexity of the matching, because the two expansion schemes (expansion of potentials in powers of momenta vs. expansion of the K matrix in powers of relativistic invariants) do not align term-by-term. We thus expect that matching would not proceed in a one-to-one manner but instead relate linear combinations of higher-order RFT couplings to specific chiral EFT coefficients.

\section{Numerical examples}
\label{sec:numerical}

In this section, we present a numerical comparison of the $DD^*$ amplitude obtained using the three approaches considered in this work: the RFT formalism of \Cref{sec:RFT}, the LS equation of \Cref{sec:LSE}, and the BRH approach of \Cref{app:RH}. We use the single-channel versions of the formalisms in all cases. We solve the full relativistic RFT equations of \Cref{sec:RFT:restricted}, keeping the full expression for $\cM_2$, i.e. without assuming pole dominance. For the LS equation, we use the NR reduced form described in \Cref{sec:LS-OPE,sec:LS-full}, in particular using $V_{\pi,S}^{\rm NR}$ for the OPE potential. For the BRH approach, we use the relativistic form of the equations, as given in \Cref{app:RH:restriction}. We choose the constants in the approaches based on the matching equations, \Cref{eq:matching-c0,eq:RFT-couplings}. Since we know from \Cref{sec:NR-RFT,app:RH-NR} that the formalisms agree after NR reduction and assuming pole dominance, we aim to see how they compare in the absence of these approximations.

\begin{figure}[t!]
    \centering
    \begin{subfigure}[b]{0.48\textwidth}
        \includegraphics[width=0.99\textwidth]{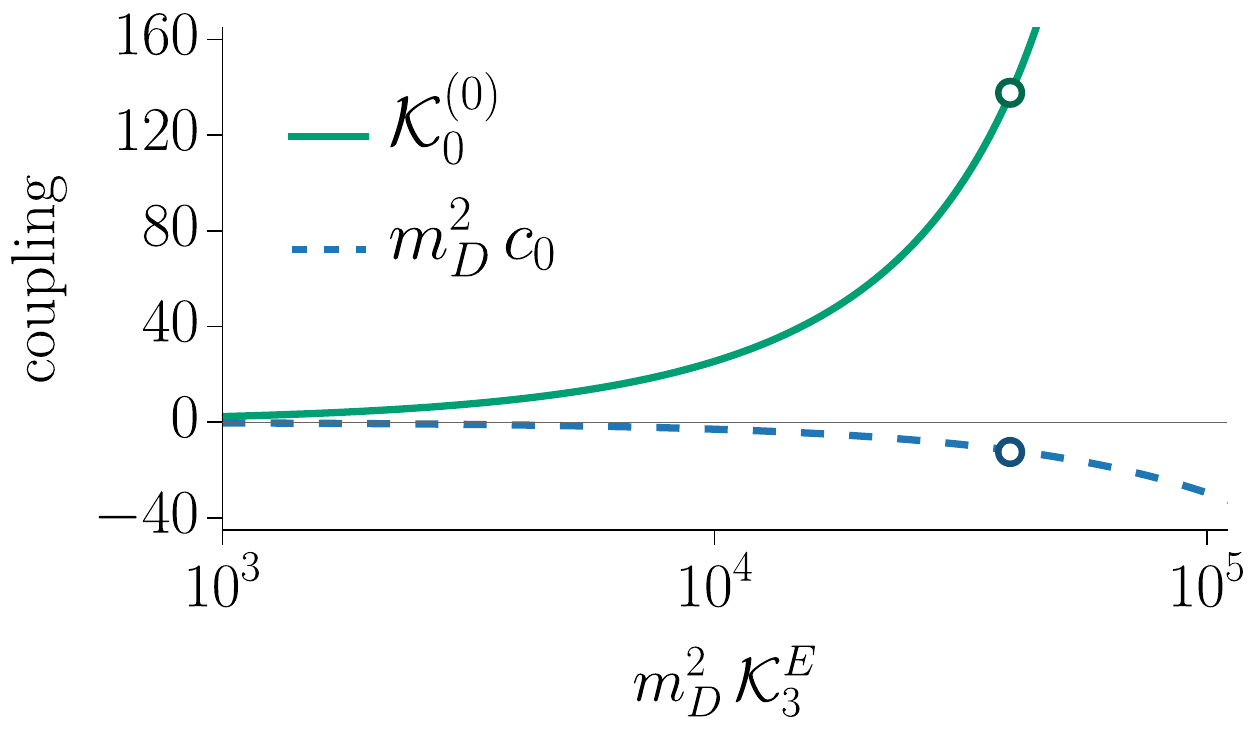}
    \end{subfigure}
    \hfill
    \begin{subfigure}[b]{0.48\textwidth}
        \includegraphics[width=0.99\textwidth]{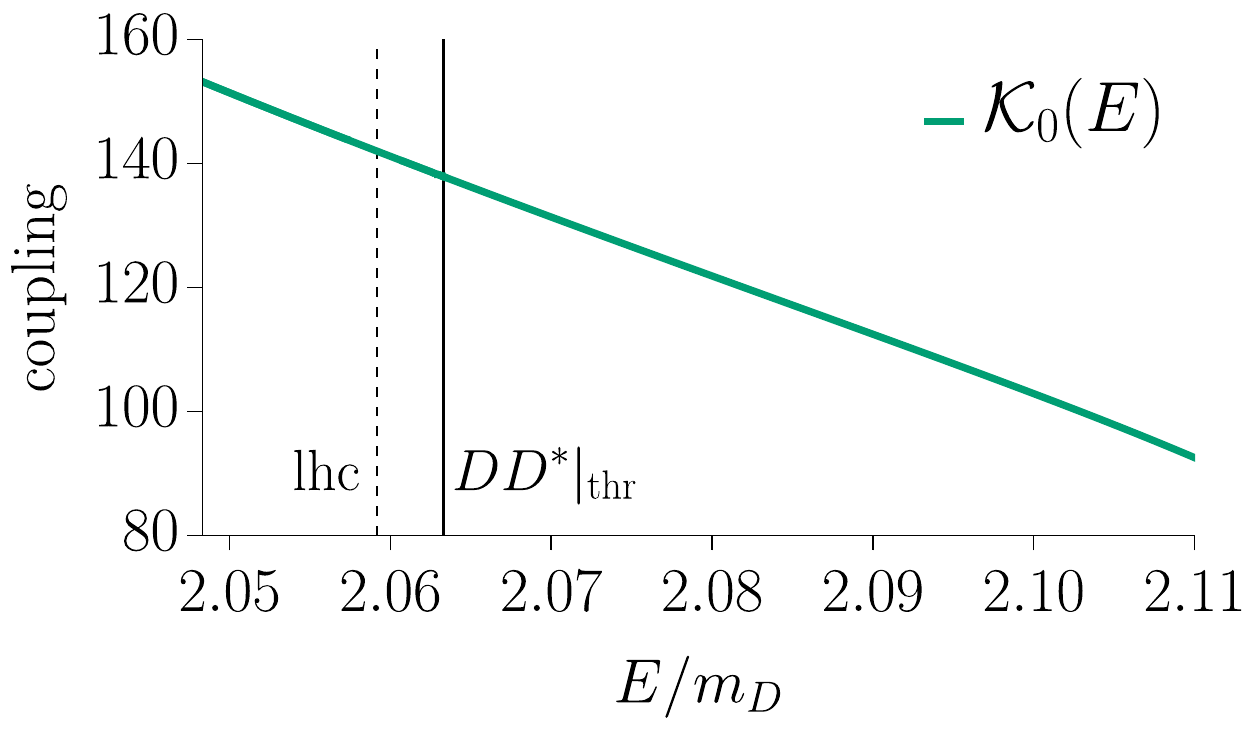}
    \end{subfigure}
    \caption{ Comparison of dimensionless couplings appearing in the RFT, LS, and BRH approaches. Left panel: dependence of the LS coupling $c_0$ and BRH coupling $\cK_0^{(0)}$ (defined in the text) on the RFT K matrix parameter $\cK_3^E$, given in~\Cref{eq:coupling-1} and~\Cref{eq:RFT-couplings}, respectively. Circled points correspond to the ``best-match'' coupling $\cK_3^E \, m_D^2 = 4 \cdot 10^4$. Right panel: energy-dependent coupling $\cK_0(E)$ in the BRH approach as a function of energy, for $\cK_3^E \, m_D^2 = 4 \cdot 10^4$. 
    }
    \label{fig:couplings}
\end{figure}

We compare formalisms using the following (unphysical) masses, 
    \begin{equation}
    m_D = 1927~ {\rm MeV} \, , \quad 
    m_{D^*} = 2049~ {\rm MeV} \, , \quad 
    m_\pi = 280~ {\rm MeV} \, , 
    \label{eq:masses}
    \end{equation}
which correspond closely to the values used in the lattice QCD calculation of  ref.~\cite{Padmanath:2022cvl}, as well as in our previous work analyzing this lattice data~\cite{Dawid:2024dgy}. As described in previous sections, we use a hard momentum cutoff of $\Lambda = 500\;$MeV for all three approaches. The other parameters that are needed are the pion decay constant, for which we use $f_\pi = 130~{\rm MeV}$, and the heavy meson ChPT coupling, which we set to $g = 0.5464$. The $D\pi$ two-body K matrix of~\Cref{eq:2b-kmatrix} is characterized by ERE parameters $r/m_D = -6.5$ and $a m_D^3 = 19.98$. These choices lead to 
    \begin{align}
    \begin{split}
    \frac{m_\pi}{m_D} = 0.1453 \, , \quad 
    \frac{\Delta}{m_D} = \ 0.06331 \, , \quad
    \frac{|q_0|}{m_D} = 0.1308 \, , \quad  
    \frac{\Lambda}{m_D} = 0.2595 \, , 
    \end{split}
    \end{align}
where we recall that $\Delta=m_{D^*}-m_D$, while $q_0$ is given by the NR expression, \Cref{eq:NRmomenta}. We use this value of $q_0$ in the matching equations, while we use the relativistic form, \Cref{eq:q0def}, when solving the RFT and BRH equations, which leads to $|q_0|/m_D=0.1266$.

Given the above choices, the residue of the pole in $\cM_2$ used in the RFT equations, \Cref{eq:zeta}, is given by $\zeta^2/m_D^2 = 2.235$ (using the relativistic value for $q_0$) and in the matching condition $\zeta^2/m_D^2 = 2.386$ (using the NR value for $q_0$). In the modified phase space, \Cref{eq:mod-rho}, we follow ref.~\cite{\DRS} and use $c_{\rm PV}/m_D^3 = -0.25$. The quantities entering the RFT to LS matching formula, \Cref{eq:matching-c0}, are then
    \begin{align}
    c_L \, m_D = 0.03133 \, \sqrt{ \cK_3^E \, m_D^2 }\, , \quad \cI_{\rm NR, 0} = -8.224 \cdot 10^{-8} (\cK_3^E m_D^2) \,,
    \end{align}
leading to the matching result
    \begin{align}
    \label{eq:coupling-1}
    c_0 \, m_D^2 &= - \frac{ a_1 \, (\cK_3^E m_D^2)}{ 1 - a_2 \, (\cK_3^E m_D^2) } \, , \quad a_1 = 2.927 \cdot 10^{-4} \, , \quad a_2 = 8.224 \cdot 10^{-8} \, , 
    \end{align}
This expression is plotted in the left panel of \Cref{fig:couplings}.  

The corresponding matching formula for the BRH coupling, \Cref{eq:RFT-couplings}, is given by
    \begin{align}
    \label{eq:num-energy-dep-K0}
    \cK_0(E) = \frac{b_1 \, \cK_3^E m_D^2}{ 1 - b_1 \, \big[ b_2 + \cI_{\rm NR,0}^{\rm BRH}(E) \big] \, \cK_3^E m_D^2 } \, , \quad 
    b_1 = 2.342 \cdot 10^{-3} \, , \quad 
    b_2 = 3.512 \cdot 10^{-5} \, .
    \end{align}
where the energy-dependent integral $\cI_{\rm NR,0}^{\rm BRH}(E)$ is given by \Cref{eq:INR0RH}. If we take the threshold value for this integral, 
$\cI_{\rm NR,0}^{\rm BRH}(E_{\rm th}) = 3.387 \cdot 10^{-3}$, we obtain the quantity denoted $\cK_0^{(0)}$, which is also plotted in the left panel of \Cref{fig:couplings}. An example of the energy dependence of $\cK_0(E)$ is shown in the right panel.

\begin{figure}[t!]
    \centering
    \includegraphics[width=0.8\textwidth]{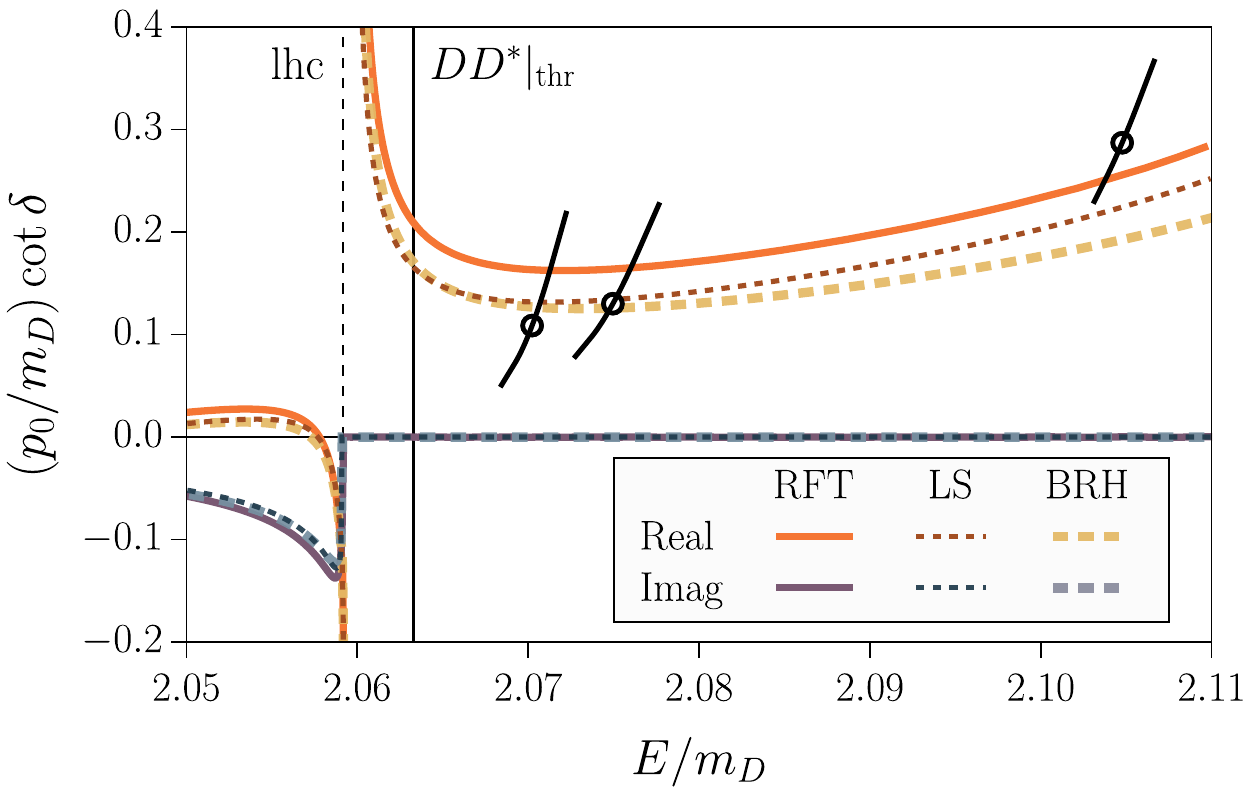}
    \caption{Comparison of $DD^*$ $s$-wave phase shifts (plotted as $p_0 \cot \delta/m_D$) obtained using the three-body RFT approach (solid lines) using $\cK_3^E m_D^2 = 4 \cdot 10^4$; the LS equation (dashed thin lines) for $c_0 m_D^2 = -11.75$; and the BRH approach  (dashed thick lines) with energy-independent constant $\cK_0^{(0)} = 137.9$. Black points are lattice QCD data from~\cite{Collins:2024sfi}. The left-hand cut in the $DD^*$ channel due to $u$-channel pion exchange is shown by the vertical dashed line.
    }
    \label{fig:phase-shifts}
\end{figure}

In~\Cref{fig:phase-shifts}, we show solutions of the integral equations in the three considered formalisms. For the RFT solution, we follow the numerical prescriptions discussed in sec.~2.2.3 of ref.~\cite{Dawid:2024dgy}; in particular, we use deformed integration contours to avoid the $D^*$ pole, as shown in fig.~1 therein. The couplings in these solutions are obtained as follows. We choose the RFT constant to be $\cK_3^E m_D^2 = 4 \cdot 10^4$, so as to provide an approximate ``best match'' to the lattice points from ref.~\cite{Collins:2024sfi}. The LS constant is then fixed by the matching condition \Cref{eq:coupling-1}, yielding $c_0 m_D^2 = -11.75$, while the BRH coupling is given by the energy independent value $\cK_0^{(0)} = 137.9$.

The figure shows reasonable matching between the three results across the entire energy range. The LS and BRH curves match closely at and below the threshold but differ with increasing energy. The RFT result is offset from those from the LS and BRH approaches for most energies.\footnote{%
We have checked numerically that, if we make the NR approximations described in \Cref{sec:RFT-NR-ladder,sec:mdf3reduce}, along with the pole-dominance approximation, the RFT results do agree with those from the LS equation---see \Cref{fig:compare-approximations} and associated text.
} 
Differences between the results are expected, as the RFT and BRH results do not use the NR approximation, and the RFT result does not assume pole dominance in $\cM_2$.

\begin{figure}[t!]
    \centering
    \includegraphics[width=0.8\textwidth]{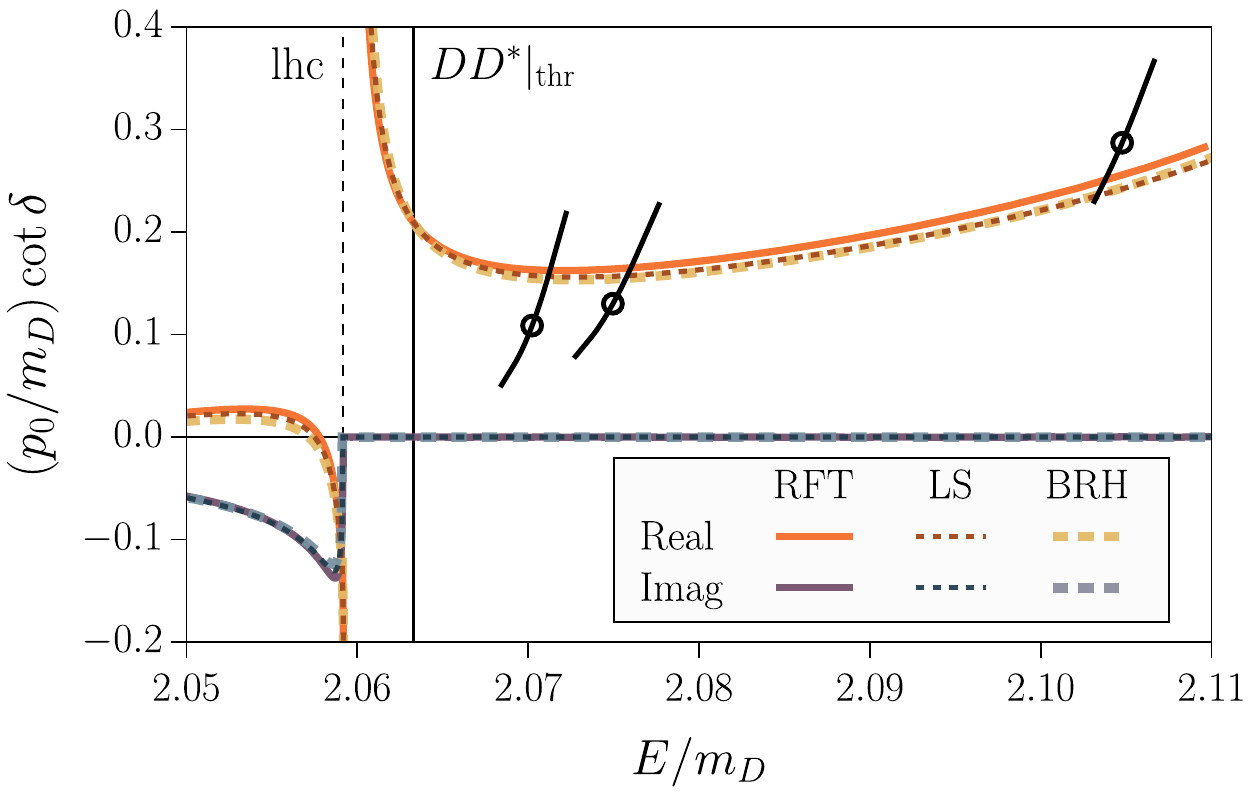}
    \caption{Phase shifts for the choice of couplings such that all three formalisms match at the threshold. We take $\cK_3^E m_D^2 = 4 \cdot 10^4$ in the three-body RFT approach (as in \Cref{fig:phase-shifts}),  $c_0 m_D^2 = -10.1$ in the LS equation, and the energy-dependent $\cK_0(E)$ corresponding to $\cK_3 m_D^2 = 3.5 \cdot 10^4$ in the BRH equation, \Cref{eq:num-energy-dep-K0}. Conventions as in~\Cref{fig:phase-shifts}.
    }
    \label{fig:phase-shifts-best-match}
\end{figure}

Additionally, we have computed the pole positions of the state corresponding to the $T_{cc}^+$ at the heavier-than-physical pion mass for parameters from~\Cref{fig:phase-shifts,fig:phase-shifts-best-match}. The pole positions are listed in~\Cref{tab:tab1} and were obtained by continuing the $DD^*$ amplitude to the unphysical Riemann sheet associated with the unitarity cut. Given the simple LO matching between models from~\Cref{fig:phase-shifts}, and the approximate ``eyeball'' matching in~\Cref{fig:phase-shifts-best-match}, we find reasonably good quantitative agreement in the pole positions across all formalisms in both cases.

\begin{table}[]
    \centering
    \begin{tabular}{c|c|c|c}
       ~  & RFT & LS & BRH \\ \hline \hline
      \Cref{fig:phase-shifts} & $2.0573 \pm 0.0079 \, i$ & $2.0587 \pm 0.0068 \, i $  & $2.0591 \pm 0.0073 \, i $ \\
      \Cref{fig:phase-shifts-best-match} & $2.0573 \pm 0.0079 \, i$ & $2.0578 \pm 0.0081 \, i $  & $2.0585 \pm 0.0082 \, i $ \\
    \end{tabular}
    \caption{Positions of the putative $T_{cc}^+$ poles in the complex energy variable, $E/m_D$, on the second sheet of the $DD^*$ amplitude. We list values corresponding to all three approaches, with model parameters fixed as in~\Cref{fig:phase-shifts,fig:phase-shifts-best-match}. }
    \label{tab:tab1}
\end{table}

We expect the matching conditions to receive corrections due to the approximations made in deriving them. To investigate this, we have adjusted the couplings in the LS and BRH solutions such that the resulting scattering amplitudes agree at the $DD^*$ threshold. The results are shown in \Cref{fig:phase-shifts-best-match}.
We find that a $\sim 15\%$ decrease in the magnitude of $c_0$ brings the LS curve into much better agreement with the RFT result over the entire energy range considered. As for the BRH result, a $\sim 17\%$ decrease in $\cK_0^{(0)}$ brings the threshold value into agreement with the RFT result. We also find that using the energy-dependent form for $\cK_0(E)$ brings the entire curve into almost exact agreement with the LS result. While this is a non-leading effect in the NR expansion, it is clearly dominant for increasing energies. Overall, we conclude that the differences between different approaches can be absorbed into shifts in the couplings to good approximation.

We include additional comparison plots at the end of~\Cref{app:RH} and~\Cref{app:offshell}, which
showcase the effect of the energy-dependent $\cK_0$ in the BRH formalism---see~\Cref{fig:phase-shifts-energy}---and the significance of the ``pole dominance'' approximation (see~\Cref{fig:compare-approximations}).

\section{Conclusions}
\label{sec:conclusions}

In this work, we have compared three methods that are used to study the $D D^*$ system, as well as many other systems of interest in hadronic physics. These methods are the well-established Lippmann-Schwinger approach, and the more recently introduced three-particle relativistic field theory and two-particle Bai\~ao-Raposo--Hansen approaches. They differ in several important respects. First, the LS equation is partially nonrelativistic, while the RFT and BRH approaches are fully relativistic. Second, off-shell potentials play a central role in the LS approach, while all amplitudes involved in the RFT approach are on shell. Third, the RFT approach treats the $D^*$ as a pole in the $D\pi$ amplitude rather than a fundamental degree of freedom.\footnote{%
We note, however, that ref.~\cite{Du:2021zzh}, which uses the LS approach, includes self-energy contributions to the $D^*$ propagator from $D\pi$ loops, as well as the effects of the $D\gamma$ decay channel, so that the $D^*$ becomes a resonance pole.
}
And, finally, the RFT approach includes interactions in subchannels that have no obvious counterpart in the two-particle approaches, specifically $s$-wave $D\pi$ and $DD$ and $d$-wave $DD^*$ contributions.\footnote{Such effects could, in principle, be incorporated in the two-body formalisms by deriving them from the full set of Faddeev equations.} It is thus of considerable interest to determine if and where they agree, and the numerical size of any differences.

We find that the RFT and LS approaches agree if we make two approximations: keeping the leading terms in the NR limit, and assuming that the $D^*$ pole term dominates the $D\pi$ amplitude in the RFT approach. In addition, agreement requires that short-distance contributions to the interactions are appropriately matched. By contrast, agreement between BRH and LS approaches requires only keeping the leading terms in the NR limit, as well as appropriately matching short-distance terms.
We stress that the agreement between the three-particle RFT and two-particle LS and BRH approaches is nontrivial, given the differences in the initial sets of equations.
Nevertheless, since the three approaches claim to provide a general framework that encompasses the same underlying physics, it is reassuring that they do agree in the stated limits.

We have also performed a numerical comparison of the three approaches, keeping the full relativistic expressions for the RFT and BRH cases, as well as including the full $D\pi$ amplitude in the former. We find small but noticeable differences among all three methods when using the matching conditions determined in the NR and pole-dominance limit. However, we demonstrate that these differences can be largely eliminated by introducing relatively small adjustments to the short-distance constants.

An important issue that is not addressed in this work is how the three approaches compare when applied in finite volume to predict the spectrum. The finite-volume counterpart of the LS approach is the use of the plane-wave basis, along with EFT interactions, following ref.~\cite{Meng:2021uhz}. This has been applied to the $D D^*$ system in refs.~\cite{Meng:2024kkp, Abolnikov:2024key, Collins:2024sfi, Prelovsek:2025vbr}. It would be very interesting to determine whether---and under which approximations---this approach is equivalent to the RFT and BRH quantization conditions. We anticipate that an important aspect of this comparison  will be the choice of cutoff function: in the RFT approach, a smooth cutoff is required in finite volume, while the plane-wave method is typically implemented with a hard cutoff. While we leave this question for future work, we speculate that differences between the finite-volume formalisms will be more pronounced, as the proximity of the $DD\pi$ threshold may lead to enhanced finite-volume effects not captured by the two-body formalisms. 

Our studies have brought to light a technical issue in the RFT approach, which arises from the use of a smooth cutoff function, which is mandatory in applications to finite-volume spectra.
The cutoff function turns out to act as a form factor suppressing the $DD^*\pi$ vertex, thus weakening the OPE contribution. This has not been studied in previous work~\cite{Hansen:2024ffk,Dawid:2024dgy}. While this change can in principle be compensated by changes in the three-particle K matrix, in \Cref{app:cutoff} we propose a cleaner solution that utilizes a class of modified---but still smooth---cutoff functions that avoid this issue altogether.

We close by mentioning another issue deserving of further study. In the LS approach, the OPE exchange contribution uses the leading order EFT form, with the coupling $g$ typically determined from the $D^*\to D\pi$ decay width. This neglects higher-order EFT contributions to the difference between an on-shell decay and the off-shell vertex in OPE. By contrast, in the RFT approach this vertex arises from the $D\pi$ on-shell scattering amplitude continued to subthreshold momenta. This quantity, which is physical, can be determined in principle from a lattice QCD study of the $D\pi$ system using the two-particle quantization condition. This has been done, albeit for unphysical quark masses, in ref.~\citep{Mohler:2012na, Becirevic:2012pf, Moir:2016srx, Gayer:2021xzv, Yan:2024yuq}. Extending this to physical masses will be a necessary future step.

\section*{Acknowledgments}

We thank Andre Bai\~ao Raposo, Vadim Baru, Raul Brice\~no, Max Hansen, and Saša Prelovšek for useful discussions. SMD and SRS acknowledge the financial support through the U.S. Department of Energy Contract No. DE-SC0011637. This work contributes to the goals of the USDOE ExoHad Topical Collaboration, contract DE-SC0023598.

\appendix

\section{Bai\~ao-Raposo--Hansen approach}
\label{app:RH}

In ref.~\cite{Raposo:2023oru}, Bai\~ao Raposo and Hansen proposed a new two-body finite-volume formalism. It resolves the left-hand cut problem by modifying the typical all-orders diagrammatic derivation of the L\"uscher-like quantization condition~\cite{Kim:2005gf}, singling out one-particle exchange diagrams contributing to the nearest discontinuity of the two-to-two amplitude below the threshold. Its extension to coupled-channel scattering of particles with non-zero spin was subsequently provided by Bai\~ao Raposo, Brice\~no, Hansen, and Jackura in ref.~\cite{Raposo:2025dkb}. It thus can be applied to the $DD^*$ system, as described in sec.~3.3.2 of ref.~\cite{Raposo:2025dkb}.

In the infinite-volume portion of the proposed formalism, the scattering amplitude is described by equations of Bethe-Salpeter form. These are structurally very similar to those considered in the three-body RFT method discussed in \Cref{sec:RFT}. In the following, we first summarize these equations, and then describe their NR reduction. This allows a comparison with the RFT and LS approaches.

\subsection{Relativistic BRH equation}
\label{app:RH:rel}

We begin by presenting an overview of the BRH approach, with the aim of showing the analogy to the RFT forms described in \Cref{sec:RFT:gen}.

In the BRH approach, one considers two-particle amplitudes in which one of the particles is placed on shell, while the other is in general off shell. For a fixed total four momentum $(E, \bm 0)$, the amplitudes thus depend on the final and initial three-momenta, $\bm p$ and $\bm k$, of the on-shell particle. After partial-wave projection, the amplitudes depend on the magnitude of these momenta, together with the orbital angular momentum, $L$, the total spin, $S$, and the total angular momentum, $J$. In general there is also an index, $a$, labeling different flavor channels. Combining these indices into a generalized matrix, we can write the master equation for the BRH $DD^*$ two-body partially off-shell amplitude as [see Eq.~(3.3) of ref.~\cite{Raposo:2025dkb}]
    \begin{align}
    \bm \cM^{\rm BRH}_{DD^*} &= \bm \cM_{\cE} + \Delta \bm \cM \,, 
    \label{eq:RHdecomp}\\
    \Delta \bm \cM &= \left( \bm 1 + i \bm \cL^{\rm BRH} \right) \left( \bm \cM_0^{-1} + \bm \cC \right)^{-1} \left(  \bm 1 + i \bm \cR^{\rm BRH} \right) \, .
    \label{eq:Delta_M}
    \end{align}
Here we use the superscript ``BRH'' where necessary to distinguish the amplitude from others appearing in this work. The multiplication of the generalized matrices is defined as~\cite{Raposo:2025dkb}
    \begin{align}
    [\bm \cA \, \bm \cB]_{a' L' S'; a L S}^J(p, k) = 
    \sum_{n} \, 
    \sum_{L''} \,
    \sum_{S''}
    \int \frac{dq \, q^2}{2\pi^2} 
    \cA_{i L'S' ; n L''S'' }^J(p, q) \, 
    \cB_{n L''S''; j LS}^J( q, k) \, .
    \label{eq:RH-mulip}
    \end{align}
While this differs from the definition of multiplication of generalized matrices in the three-body RFT approach [see \Cref{eq:3-multip}], our use of the same notation for the two cases should not lead to confusion.

The decomposition of \Cref{eq:RHdecomp} is analogous to that of \Cref{eq:M3} in the RFT formalism. In particular, the amplitude $\bm \cM_{\cE}$ can be considered as a two-body version of the pair-spectator ladder amplitude $\bm \cD$. It is a solution of a Bethe-Salpeter-like equation with the integration kernel truncated to include only the one-particle exchange $t$- and $u$-channel diagrams and with one of the particles in the two-body intermediate state put on-shell [see Fig.~5 of ref.~\cite{Raposo:2025dkb}]. This equation takes the form
    \begin{align}
    \bm \cM_{\cE} = \bm \cE - \bm \cE \, \bm \Delta_2 \, \bm \cM_{\cE} \, ,
    \label{eq:RH-ladder}
    \end{align}
where $\bm \cE$ represents a sum of $t$- and $u$-channel one-particle exchange diagrams, and $\bm \Delta_2$ is a two-particle propagator.

We now turn to $\Delta \cM$, the second contribution in the decomposition \Cref{eq:RHdecomp}. $\Delta \cM$ is the analog of $\bm \cM_{\rm df,3}$ in the three-body formalism, and is given in \Cref{eq:Delta_M}. Here the amplitude $\bm \cM_0$ is defined as the solution of the full Bethe-Salpeter equation with $\cE$ removed from the kernel. In other words, $\bm \cM_0$ would fully describe the scattering process if there were no one-particle exchanges, and thus parametrizes all short-range interactions that do not contain contributions from OPE processes. The endcap functions in \Cref{eq:Delta_M} describe final state rescatterings due to the iterated one-particle exchanges,
and are given by
    \begin{align}
    i \bm \cR^{\rm BRH} & = - \bm Q \, \bm \Delta_2 \, \bm \cM_{\cE} \, , \quad  i \bm \cL^{\rm BRH} = - \bm \cM_{\cE} \, \bm \Delta_2 \, \bm Q \, , 
    \label{eq:RHendcap}
    \end{align}
 where $\bm Q$ are angular-momentum-dependent threshold suppression factors. 
 The final quantity in \Cref{eq:Delta_M} is $\cC$, which is an energy-dependent short-range contribution to the amplitude that is generated by interactions in which one-particle exchanges appear in intermediate states,
    \begin{align}
    \bm \cC = - \bm Q \, \bm \Delta_2 \, \bm \cM_{\cE} \, \bm \Delta_2 \bm Q \, .
    \end{align}

General expressions for $\bm Q$, $\bm \Delta_2$ and $\bm \cM_0$ can be found in ref.~\cite{Raposo:2025dkb}, and are not repeated here.
We now turn to the single-channel restriction of this formalism,
in which we do give explicit expressions for all quantities.

\subsection{Single-channel restriction of the BRH equation}
\label{app:RH:restriction}

As in the RFT analysis of \Cref{sec:RFT:restricted}, we now assume that only the $(L,S)=(0,1)$ channel (with $J^P=1^+$) contributes. We drop the superscript denoting $J$ and the $(LS)$ subscripts. The partial-wave projected BRH ladder equation,~\Cref{eq:RH-ladder}, then reduces to the following form (see eq.~(3.23) of ref~\cite{Raposo:2025dkb}),
    \begin{align}
    \cM_{\cE}(p,k) = \cE(p,k) - \int \frac{dq \, q^2}{2\pi^2} \, \cE(p,q) \, \Delta_2(q; E) \, \cM_{\cE}(q,k) \, ,
    \label{eq:Meps}
    \end{align}
where we note that only the magnitudes of the external momenta enter. Here, following sec.~3.3.2 of ref.~\cite{Raposo:2025dkb}, the partial-wave-projected one-particle exchange amplitude is
    \begin{align}
    \cE(p,k) = \cC_1 + \cC_2 \, Q_0(z')  \, ,
    \label{eq:ERH}
    \end{align}
where
    \begin{align}
    z' &= \frac{1}{2pk} \left( m_D^2 + m_{D^*}^2 - m_\pi^2 - 2 \, \omega_D(p) \, \omega_{D^*}(k) \right) \, ,
    \end{align}
and the coefficients are given by
    \begin{align}
    \cC_1 & = - \frac{g'^2}{6} \Big[ \frac{2}{3} (1 - \gamma_1) (1 - \gamma_3) - \gamma_1 \gamma_3 + z' f - f' \Big] \, , 
    \label{eq:C1}
    \\
    \cC_2 & = - \frac{g'^2}{6} \Big[(1-z'^2) f + z' f' \Big] \, ,
    \label{eq:C2}
    \\
    f &= \gamma_3 \left( \frac{\beta_3}{\beta_2} + z' \right) 
    + \gamma_1 \left( \frac{\beta_1}{\beta_4} + z' \right) 
    - z'  \, , \\
    f' &= \gamma_1 \gamma_3 z' \left( \frac{\beta_3}{\beta_2} + \frac{\beta_1}{\beta_4} + z' \right) 
    + \frac{\beta_3 \beta_1}{ \beta_2 \beta_4} \gamma_3 \gamma_1  \,,
    \end{align}
with,
    \begin{align}
    \gamma_i = \frac{1}{\sqrt{1 - \beta_i^2}} \, , \quad 
    \beta_{1} = \frac{k}{\omega_D(k)} \, , ~ 
    \beta_{2} = \frac{k}{\omega_{D^*}(k)} \, , ~ 
    \beta_{3} = \frac{p}{\omega_{D}(p)} \, , ~ 
    \beta_{4} = \frac{p}{\omega_{D^*}(p)} \, .
    \end{align}
Finally, $g'$ is the coupling constant of the $DD^*\pi$ vertex in the BRH normalization.
    
The remaining quantity in \Cref{eq:Meps} is the Green function,
    \begin{align}
    \Delta_2(q; E) = \frac{\omega_D(p_0)}{\omega_D(q)} \frac{H(q)}{2 E (p_0^2 - q^2 + i\epsilon )} \, ,
    \end{align}
where we recall that $p_0$, defined in \Cref{eq:p0def}, is the momentum of the $D$ and $D^*$ in the absence of interactions, $E = \omega_D(p_0)+\omega_{D^*}(p_0)$. The smooth cutoff function $H(q)$ is in the same class as that appearing in the three-body RFT formalism, e.g., in \Cref{eq:G0101}.

We now turn to the second part of the BRH amplitude, namely the quantity denoted $\Delta \cM$ in Eq.~\eqref{eq:Delta_M}. Its partial-wave projection in the restricted system is given by
    \begin{align}
    \Delta \cM(p,k) = \left[
     1 -  \cL^{\rm BRH}(p) \right] \,
    \frac{1}{\cM^{-1}_0 + \cC } \,
    \left[ 1 - \cR^{\rm BRH}(k) \right] \,,
    \label{eq:DeltaMrestricted}
    \end{align}
where the endcap functions are
    \begin{align}
    \cL^{\rm BRH}(p) &= \int \frac{dq \, q^2}{2\pi^2} \,
    \cM_{\cE}(p,q) \, 
    \Delta_2(q, E) \, ,
    \label{eq:LRHrestricted}
    \\
    \cR^{\rm BRH}(k) &= 
    \int \frac{dq \, q^2}{2\pi^2} \, \Delta_2(q; E) \, \cM_{\cE}(q,k) \, , 
    \label{eq:RRHrestricted}
    \end{align}
and the energy-dependent constant is
    \begin{align}
    \cC = - \int \frac{dr \, r^2}{2\pi^2} 
    \int \frac{dq \, q^2}{2\pi^2} \, \Delta_2(r; E) \, \cM_{\cE}(r,q) \, \Delta_2(q, E) \, .
    \label{eq:Crestricted}
    \end{align}
Finally, $\cM_0$, the amplitude arising from short-range interactions, can be written in terms of a real $K$ matrix and the $s$-channel unitarity cut,
    \begin{equation}
    \cM_0^{-1} = \cK_0^{-1} - i \rho\,,\qquad
    \rho = \frac{p_0}{8 \pi E} \,.
    \end{equation}
$\cK_0$ is similar to the standard two-particle K matrix, except that it has no LHC due to OPE. It is analogous to the divergence-free three-body K matrix $\bm \cK_3$ in the three-particle formalism [see \Cref{eq:LTR,eq:T}]. It can be parametrized using an effective-range expansion.

There are no barrier factors in the expressions for $\Delta \cM$, since we are considering only the $\ell=0$ channel, for which they are absent. We also stress that, to obtain the physical scattering amplitude, one must go on shell by setting both the external momenta $p$ and $k$ to $p_0$.

\subsection{NR limit of BRH equation and relation to RFT and LS results}
\label{app:RH-NR}

We now take the NR limit of the single-channel BRH equations, using the same power-counting as in the RFT case, \Cref{eq:power-counting2}. Unlike in the RFT analysis of \Cref{sec:NR-RFT}, here we do not need to invoke the pole-dominance approximation.

We begin with the ladder amplitude. We find that $\cE$, \Cref{eq:ERH}, reduces to
    \begin{align}
    \cE_{\rm NR}(p,k) = - \frac{g'^2}{3} \left( \frac{p^2 + k^2}{2 p k } Q_0(z_{\rm NR}) + Q_1(z_{\rm NR} )\right) = - \frac{g'^2 q_0^2}{3} \, G_{\rm NR}(p,k) \, , 
    \label{eq:ERHNR}
    \end{align}
where $z_{\rm NR}$ is given in~\Cref{eq:zNR}, and we used the relation $Q_1(x) = x \, Q_0(x) - 1$ between Legendre functions.
The second equality, which follows from the definition of $G_{\rm NR}$, \Cref{eq:GNR}, shows the proportionality of the BRH and RFT OPE contributions in the NR limit. We also need the relation between $DD^*\pi$ coupling constants. Taking into account the isospin factor for the $DD^*$ scattering, 
we find
    \begin{align}
    g'^2 = \frac{3}{2} \, g^2_{DD^*\pi} \, .
    \end{align}
The NR approximation of the Green function is
    \begin{align}
    \Delta_2^{\rm NR}(q; E) = \frac{1}{4 m_D^2} \, \frac{1}{E_{\rm NR} - \frac{q^2}{2\mu} + i\epsilon } = \frac{1}{4 m_D^2} \, \cG(q;E) \, ,
    \end{align}
where $\cG(q;E)$ is the NR resolvent defined in \Cref{eq:Gbar}. Here, we have used $E_{\rm NR} = p_0^2/2\mu$, and replaced the cutoff function with an (implicit) hard cutoff.

Combining these results, we find that integral equation for $\cM_\cE$, \Cref{eq:Meps}, reduces to
    \begin{align}
    \cM_{\cE, \rm NR}(p,k) = \zeta^2 G_{\rm NR}(p,k) 
    - \frac{1}{4 m_D^2} 
    \int_0^{\Lambda} 
    \frac{dq \, q^2}{2\pi^2} \, 
    \zeta^2 G_{\rm NR}(p,q) \, 
    \cG(q;E) \, 
    \cM_{\cE, \rm NR}(q,k) \, .
    \end{align}
This equation is identical to that obtained from the LS equation, \Cref{eq:TSbarfinal}, and from the NR and pole-dominance limit of the RFT approach, Eq.~\eqref{eq:NRladder}, implying the equality of ladder amplitudes in that limit,
    \begin{equation}
    \cM_{\cE, \rm NR}(p,k)  = \overline T_{\pi,S}(p,k;E) 
    = d_{\rm NR}(p,k)\,.
    \end{equation}

Turning to the remainder of the amplitude, $\Delta M$,
in the NR limit \Cref{eq:DeltaMrestricted} becomes
    \begin{align}
    \Delta \cM_{\rm NR}(p,k) = 
    \left[ 1 - \cL_{\rm NR}^{\rm BRH}(p) \right] \,
    \frac{\cK_0}{1 + \cK_0( - i \rho_{\rm NR} + \cC_{\rm NR}) } \,
    \left[ 1 - \cL_{\rm NR}^{\rm BRH}(k) \right] \, .
    \label{eq:RH-M-NR}
    \end{align}
where
    \begin{align}
    \cL_{\rm NR}^{\rm BRH}(k) &= \frac{1}{4 m_D^2} 
    \int_0^\Lambda 
    \frac{dq \, q^2}{2\pi^2} \,
    \cM_{\cE, \rm NR}(k,q) \, 
    \cG(q, E) \,,
    \\
    \cC_{\rm NR} &= - \frac{1}{(4 m_D^2)^2} 
    \int_0^\Lambda \frac{dr \, r^2}{2\pi^2}
    \int_0^\Lambda \frac{dq \, q^2}{2\pi^2}  \, 
    \cG(r; E) \, 
    \cM_{\cE,\rm NR}(r,q) \, 
    \cG(q, E) \,,
    \end{align}
and $\rho_{\rm NR}$ is the NR limit of the phase space term, for which a
useful expression is
    \begin{align}
    i \rho_{\rm NR} = 
    \frac{1}{4 m_D^2} {\rm p.v.} \!
    \int_0^{\Lambda} \frac{dq \, q^2}{2 \pi^2} \, \cG(q,E) - 
    \frac{1}{4 m_D^2} \int_0^{\Lambda} 
    \frac{dq \, q^2}{2 \pi^2} \, 
    \cG(q,E) \, ,
    \label{eq:rhoNR}
    \end{align}
where p.v. indicates taking the principal value of the integral over the pole in $\cG$.

We wish to compare these expressions to those from the combined NR and pole-dominance limit of the RFT approach, which are summarized in \Cref{eq:mdfNR}. To do so we must first go on shell by setting $p=k=p_0^{\rm NR}$. Then we note that
    \begin{equation}
     \left[ 1 - \cL_{\rm NR}^{\rm BRH}(p_0^{\rm NR}) \right]
    = L_{\rm NR} (p_0^{\rm NR})\,,
    \end{equation}
so that the endcap factors match. To achieve agreement for the remaining part of $\Delta \cM_{\rm NR}$ we use \Cref{eq:rhoNR} to obtain
    \begin{align}
    \cI_{\rm NR}^{\rm BRH} &\equiv 
    \cK_0 (- i \rho_{\rm NR} + \cC_{\rm NR})  
    \\
    &= 
    \cK_0 \, \cI_{\rm NR, 0}^{\rm BRH} +
    \frac{\cK_0}{4 m_D^2} \int_0^{\Lambda} 
    \frac{dp \, p^2}{2 \pi^2} \, 
    \cG(p,E)  + \cK_0 \cC_{\rm NR} \, ,
    \end{align}
where
    \begin{align}
    &  \cI_{\rm NR,0}^{\rm BRH}(E) = 
    - \frac{1}{4 m_D^2} {\rm p.v.} 
    \int_0^{\Lambda} \frac{dp \, p^2}{2 \pi^2} \, 
    \cG(p,E) \,
    = 
    \frac{\mu \Lambda }{4 \pi^2 m_D^2} 
    \left[ 1 - x \, {\rm tanh}^{-1} \left(x \right) \right] \, , 
    \label{eq:INR0RH}
    \end{align}
where $x = \sqrt{2 \mu E_{\rm NR}}/\Lambda$. Note that this is different from the corresponding RFT quantity, $\cI_{\rm NR,0}$ in Eq.~\eqref{eq:cINR0}. 

It is now straightforward to see that $\Delta \cM_{\rm NR}(p_0^{\rm NR},p_0^{\rm NR})$ is equal to the corresponding RFT quantity, $-\zeta^2 m_{\rm df,3}(p_0^{\rm NR},p_0^{\rm NR})$, as long as the following relation holds,
    \begin{align}
    \cK_0(E) = \frac{(\zeta c_L)^2}{ 1 + \cI_{\rm NR, 0} - (\zeta c_L)^2 \, \cI_{\rm NR, 0}^{\rm BRH}(E) } \, .
    \label{eq:RFT-couplings}
    \end{align}
Note that $\cK_0$ defined this way is energy-dependent. 

\begin{figure}[t!]
    \centering
    \includegraphics[width=0.8\textwidth]{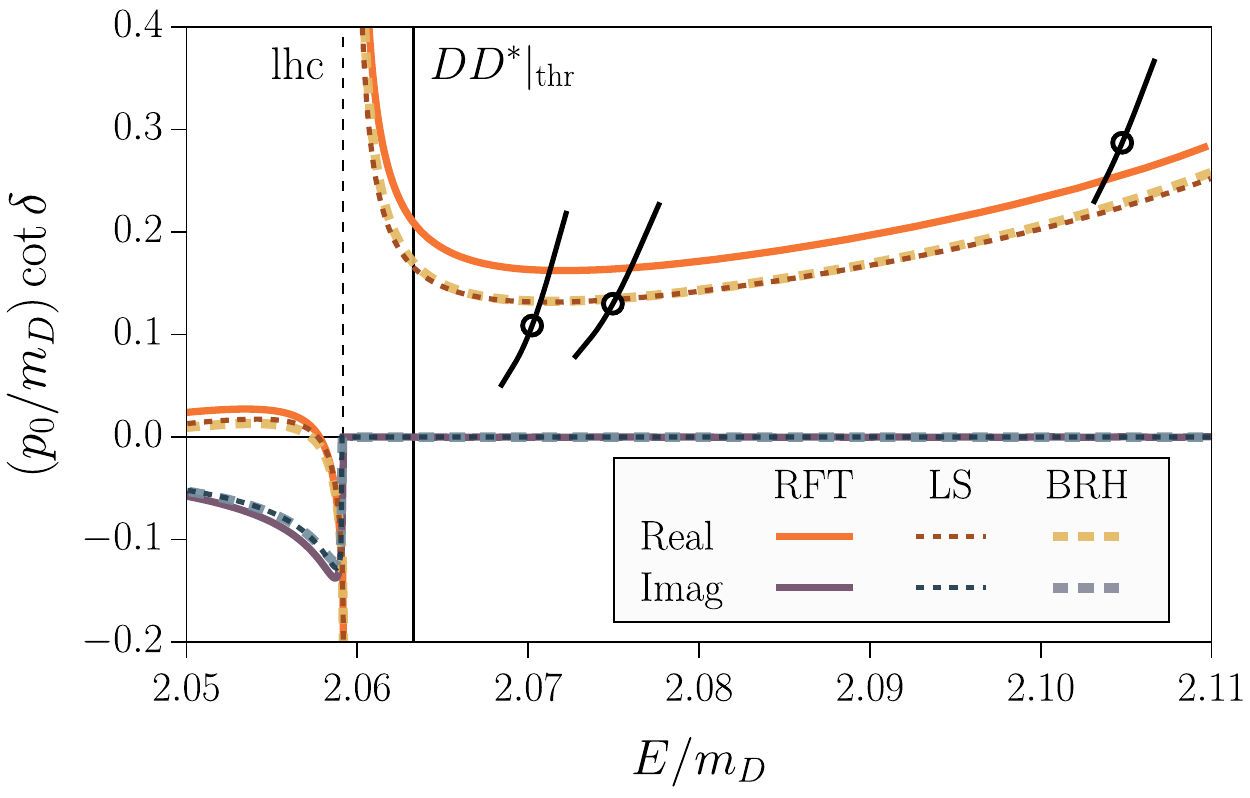}
    \caption{As in ~\Cref{fig:phase-shifts} but with the BRH result recomputed with the energy-dependent $\cK_0$ corresponding to $\cK_3^E m_D^2 = 4 \cdot 10^4$.
    }
    \label{fig:phase-shifts-energy}
\end{figure}

To match the couplings in the leading order NR approximation, we set $E=E_{\rm th}=m_D+m_{D^*}$ (equivalent to $E_{\rm NR} = 0$) in the argument of $\cI_{\rm NR,0}^{\rm BRH}$, obtaining
    \begin{align}
    \cI_{\rm NR,0}^{\rm BRH}(E_{\rm th}) = 
    \frac{\mu \Lambda }{4 \pi ^2 m_D^2} \, .
    \label{eq:LOINR0RH}
    \end{align}
Using this value in \Cref{eq:RFT-couplings} defines $\cK_0^{(0)} \equiv \cK_0(E_{\rm th})$. In~\Cref{fig:phase-shifts-energy} we show how the solution of the BRH formalism from ~\Cref{fig:phase-shifts} behaves when the leading-order $\cK_0^{(0)}$ is replaced by the energy-dependent coupling $\cK_0(E)$ given in~\Cref{eq:num-energy-dep-K0}. We see that the inclusion of the energy dependence predicted by~\Cref{eq:RFT-couplings} corrects the high-energy behavior of the BRH result, leading to a nearly-perfect agreement with the solution of the chiral EFT LS equation.

Given the agreement between RFT and LS results described in the main text, this completes the demonstration of agreement between all three approaches in the combined NR and pole-dominance limits. We stress, however, that agreement between BRH and LS equations requires only the NR limit to be taken

\section{Comments on cutoff functions}
\label{app:cutoff}

In this appendix, we discuss various choices of cutoff function in the RFT approach, as well as their relation to the hard cutoff used in the main text.

As noted in the main text, the function $H(p)$, appearing in~\Cref{eq:G0101,eq:mod-rho,eq:GNR}, regularizes momentum integrals in the RFT equations. This function is introduced in the derivation of the finite-volume quantization condition, and must be smooth (infinitely differentiable) to avoid introducing power-law finite-volume dependence~\cite{\HSQCa}. Although some of its features are general, the precise form of $H(p)$ depends on the two-particle subchannel to which it is being applied. Here we focus on the $D\pi$ channel, since this enters the restricted formalism that is the focus of the main text.

Apart from smoothness, $H(p)$ must satisfy two other requirements. First, $H(p)$ must equal unity for values of $p$ such that the two-particle invariant mass lies at or above the threshold value, $\sigma_{\rm th}(p) = (m_D + m_\pi)^2$.\footnote{%
Strictly speaking, $H(p)$ must equal unity for a small range of invariant masses below the threshold value~\cite{\BSnondegen}.
In practice, having a form that lies very close to unity in this range is sufficient.}
Second, $H(p)$ must vanish for values of $p$ such that the invariant mass lies at, or below, the nearest left-hand singularity. For the $D\pi$ system, this singularity involves two-pion exchange in the $t$ channel and occurs at $\sigma_{\rm min}(p) = m_D^2 - m_\pi^2$. The specific form of $H(p)$ used in ref.~\cite{\DRS} is
    \begin{align}
    H(p) &= J(x[\sigma(p)])\,, 
    \qquad x[\sigma] = 
    \frac{\sigma - \sigma_{\rm min}}{\sigma_{\rm th} - \sigma_{\rm min}}
    \label{eq:Hdef}
    \\
    J(x) & =
    \begin{cases}
    0 \,, & x \le 0 \, , \\
    \exp \left( - \frac{1}{x} \exp \left [-\frac{1}{1-x} \right] \right ) \,, 
    & 0<x < 1 \, , \\
    1 \,, & 1\le x \,.
    \end{cases}
    \label{eq:cutoff-J}
    \end{align}
We note, however, that there is considerable arbitrariness in the form of $J(x)$, a freedom that we will make use of in the following.

While the finite-volume part of the RFT formalism requires a smooth cutoff, the infinite-volume part does not. In particular, the integral equations satisfied by the three-particle amplitude $\cM_3$ remain valid if $H(p)$ is replaced by a hard, subthreshold cutoff. In the main text, we use such a hard cutoff to facilitate comparison with the implementation of the LS approach given in ref.~\cite{Collins:2024sfi}.

We now come to the first of two issues that we wish to address in this appendix. Since $\cM_3$ is a physical quantity, and thus independent of the choice of cutoff function, it must be that changes in cutoff function can be absorbed into changes in the only unphysical quantity in the formalism, the three-particle K matrix $\cK_3$. This holds exactly only if we allow $\cK_3$ to be an arbitrary function of kinematical variables, aside from the constraints of Lorentz invariance and appropriate particle interchange symmetries. However, we use an approximate form for $\cK_3$ depending on a single parameter, $\cK_3^E$---see \Cref{eq:K3def}. Thus the change in cutoff function will not be exactly compensated by a change in $\cK_3$. The question we want to address is how well this compensation works in practice.

\begin{figure}[t!]
    \centering
    \includegraphics[width=0.8\textwidth]{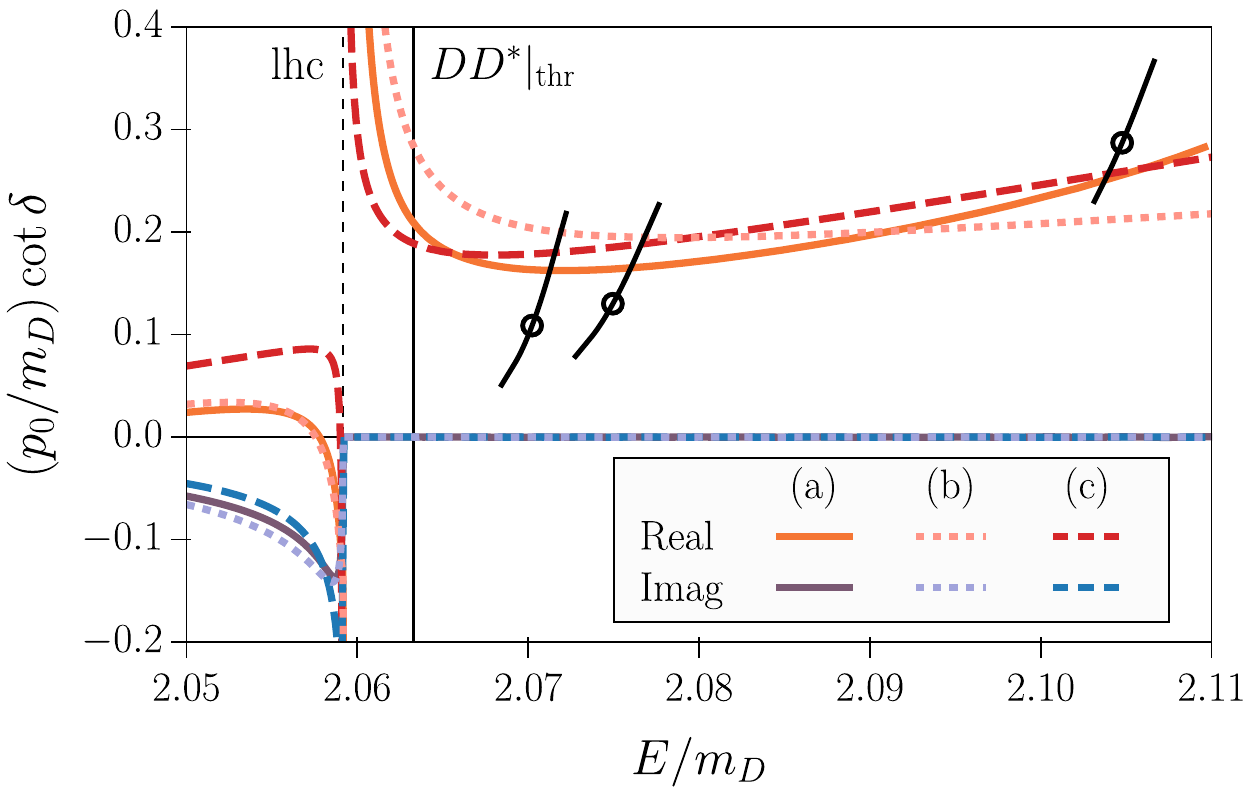}
    \caption{Comparison of $DD^*$ $s$-wave phase shifts obtained using the three-body RFT approach with: \textbf{(a)} hard cutoff for $\cK_3^E m_D^2 = 4 \cdot 10^4$, \textbf{(b)} ``bumpy'' cutoff of \Cref{eq:bumpyJ} with $\cK_3^E m_D^2 = 6 \cdot 10^4$ and $a = 10$, and \textbf{(c)} the smooth cutoff of \Cref{eq:Hdef} with $\cK_3^E m_D^2 = 2 \cdot 10^5$.
    Conventions as in~\Cref{fig:phase-shifts}.
    }
    \label{fig:compare-cutoffs}
\end{figure}

To this end, we compare the solutions of the restricted single-channel system obtained in ref.~\cite{\DRS}, using the smooth cutoff function described above, to those presented in \Cref{sec:numerical}, which use a hard cutoff. This comparison is shown in \Cref{fig:compare-cutoffs}. Solid curves labeled (a) use the hard cutoff with $\cK_3^{E,\rm hard}=4 \cdot 10^4$, and are identical to the corresponding curves in \Cref{fig:phase-shifts}. Dashed curves labeled (c) use the cutoff of \Cref{eq:Hdef,eq:cutoff-J} with $\cK_3^{E,\rm hard}=2 \cdot 10^5$. The latter value has been adjusted so that the curves approximately match each other in the energy range shown. We see that reasonable agreement can be obtained, and that this requires a significant change in $\cK_3^E$---a factor of five increase. To obtain closer agreement presumably requires the addition of higher order terms in the threshold expansion of $\cK_3$.

The second issue that we address here concerns the fact that the smooth cutoff function is smaller than unity when the invariant mass squared of the pair is $\sigma_p = m_{D^*}^2$. As noted in footnote 8 of ref.~\cite{\DRS}, this implies that the cutoff function is acting as an effective form factor for the $D^* \to D\pi$ vertex. When matching with the LS approach, as in \Cref{sec:NR-RFT}, this results in an OPE potential with a coupling smaller than that arising from leading order heavy-light meson ChPT. The reduction is by a factor of $H(p_0)^2$ [with $p_0$ given in \Cref{eq:p0def}], with the square of $H$ appearing since there are two vertices in an OPE process. This effect is numerically significant with the cutoff function used in ref.~\cite{\DRS}, $H(p_0)^2 \approx 0.5$.

Although this effect is absent for the hard cutoff used in the main text, it is relevant when using a value of $\cK_3^E$ obtained from the finite-volume quantization condition, for which a smooth cutoff is mandatory. The reduction in the OPE coefficient implies---assuming that the leading order ChPT result for the $DD^*\pi$ coupling is accurate---that the remaining part of the pion exchange contribution would have to be included in $\cK_3$ in the form of a left-hand cut with discontinuity adjusted such that the $\cM_{\rm df,3}$ contribution to $\cM_3$, after LSZ reduction, recovers the ``missing'' part of the left-hand cut in the $DD^*$ amplitude. This is not a fundamental obstacle since unitarity of the amplitude requires $\cK_3$ to be divergence-free only in the kinematical region for physical three-body scattering and leaves left-hand singularities of the three-body K matrix unconstrained. Nevertheless, it does present a practical problem when we use a smooth, truncated form for $\cK_3$, such as that of \Cref{eq:K3def}, since such a form cannot capture the missing left-hand cut contribution. Furthermore, it is contrary to the spirit of the RFT approach to the left-hand cut problem, in which the inclusion of the effects of the cut arises from OPE rather than from $\cK_3$.

A simple remedy is to have the smooth cutoff function equal unity exactly at the bound state pole, making use of the above-noted freedom in the function $J(x)$. This can be accomplished in several ways, while remaining in the class of smooth cutoff functions. One choice is to modify $J(x) \to J_b(x)$ by the inclusion of a ``bump'' at $\sigma_p = m_{D^*}^2$, such that $H_1(p)=1$ exactly at the bound state pole. This can be achieved, for example, by replacing the form \Cref{eq:cutoff-J} with\footnote{%
With $f(x)=1$ this is the ``symmetric'' form of the cutoff function introduced in ref.~\cite{Baeza-Ballesteros:2023ljl}.
}
%
    \begin{align}
    J_b(x) &=  \frac{1}{1 + f(x) \exp\left[1/x - 1/(1-x)\right]} \,,
    \label{eq:bumpyJ}
    \end{align}
where $f(x)$ is a positive function with a zero at the pole position,
    \begin{equation}
    f(x_0) = 0\,, \qquad
    x_0 = \frac{m_{D^*}^2 - \sigma_{\rm min}}{\sigma_{\rm th} - \sigma_{\rm min}}\,.
    \label{eq:x0def}
    \end{equation}
We have used a simple example of such a function,
    \begin{equation}
    f(x) = a (x-x_0)^2\,, 
    \label{eq:bumpyf}
    \end{equation}
with $a$ a parameter determining the width of the ``bump'' in $J_b(x)$.

\begin{figure}[t!]
    \centering
    \includegraphics[width=0.8\textwidth]{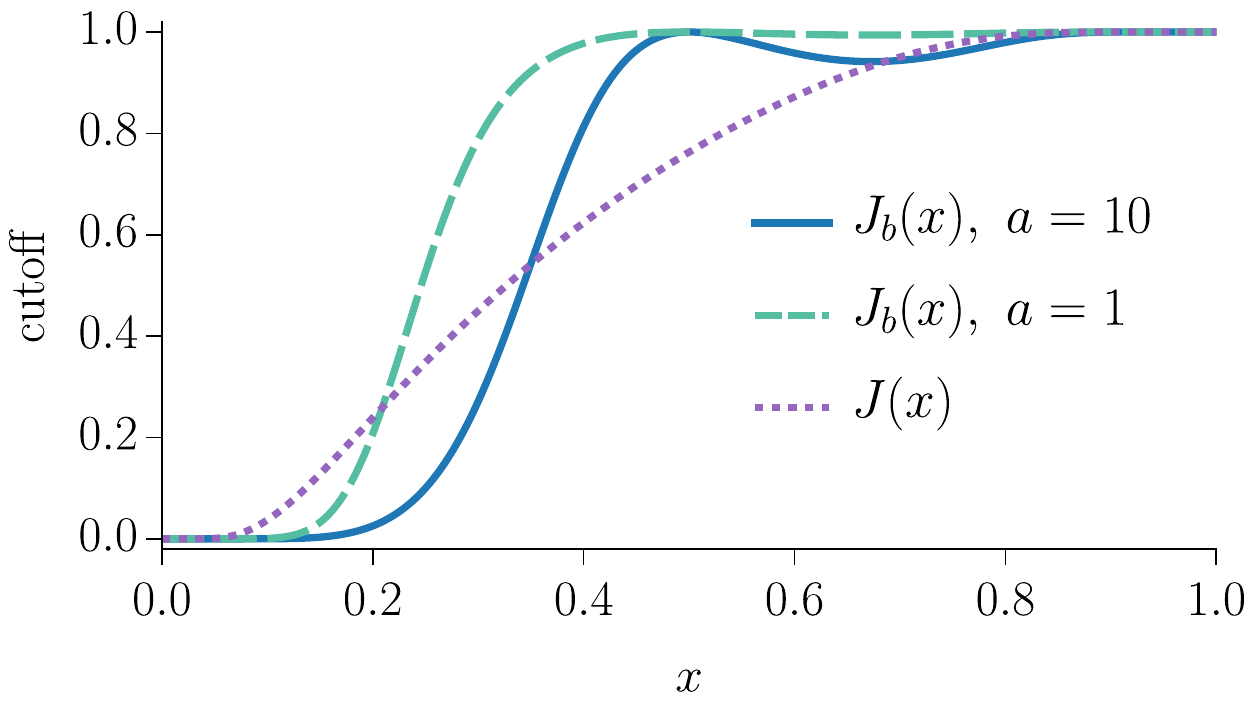}
    \caption{Comparison of different choices of the function $J(x)$ discussed in the text: $J(x)$ is defined in \Cref{eq:cutoff-J}, while the ``bumpy'' cutoff is given in \Cref{eq:bumpyf}. We use the masses of \Cref{eq:masses}, for which the position of the bump is $x_0=0.456$.
        } 
    \label{fig:compare-J}
\end{figure}

The various choices of the function $J(x)$ are compared in \Cref{fig:compare-J}. The dotted line shows the standard RFT cutoff function, which drops well below unity at the position of the $D^*$ pole, $x_0=0.456$. Two choices of ``bumpy'' cutoff, with $a=1$ and $10$, are also shown, the former showing no bump to the naked eye, while the latter having a clear bump. The advantage of the lower value of $a$ is that higher-order terms in $\cK_3$ will not be needed to fill in the artificially introduced dip present for $a=10$. The advantage of the higher value is that the width of the transition region between zero and unity is slightly larger---this region shrinks to zero as $a\to 0$. This is important because the width of the transition region introduces a scale $\Delta \sigma $ that enters into exponentially-suppressed finite-volume effects as $\exp(-\sqrt{\Delta\sigma} L)$~\cite{\HSQCa,\BHSnum}. A transition region of $\Delta x \approx 0.2$ (approximately the value for the $a=10$ curve) corresponds to 
    \begin{equation}
    \Delta \sigma \approx 0.2 (\sigma_{\rm th} - \sigma_{\rm min}) = 0.2 \times 2 m_\pi (m_D + m_\pi) \approx (1.8 m_\pi)^2
    \end{equation}
for the masses of \Cref{eq:masses}. This is an acceptably small width, given that exponentially suppressed effects proportional to $\exp(-m_\pi L)$ are neglected in RFT finite-volume analysis. In fact, the $\Delta \sigma$ for $a=1$ curve also appears acceptable.

The effect on the $DD^*$ amplitude of changing to the bumpy cutoff function can be seen from the dotted curves labeled (b) in \Cref{fig:compare-cutoffs}, for which we use $a=10$. Again we have adjusted $\cK_3^E$ by hand to attain an approximate matching---here this requires $\cK_3^E \approx 6 \cdot 10^4$. In this case, we see a more substantial change in the behavior of the amplitude above the threshold, although the behavior below the left-hand cut is very similar to that of the hard cutoff. This conclusion holds also for $a=1$, for which the result is similar to that with $a=10$ and is not shown.

It is interesting to compare the smooth cutoff functions of \Cref{fig:compare-J} to the hard cutoff at $\Lambda=500\;$MeV used in the main text. 
For a given energy, the momentum range $0\le p \le \Lambda$ corresponds to an allowed region of $x$ in the figure.
For our standard range $E/m_D=2.05-2.11$,
the minimum value of $x$ (corresponding to $p=\Lambda$) ranges almost linearly from 
$-0.04$ to $0.34$. 
Thus the hard cutoff sweeps through the region in which the soft cutoff has its main variation.\footnote{
Alternatively, the maximum value of momentum allowed by the smooth cutoff depends on energy. For our standard energy range, this maximum value changes from $q_{\rm max}/m_D = 0.25$ to $0.35$, to be compared to $\Lambda/m_D \approx 0.26$.
}
Given this substantial energy dependence, the differences in $DD^*$ phase shifts seen in \Cref{fig:compare-cutoffs} are not surprising.

\section{Effect of including nonpole term in the two-body amplitude}
\label{app:offshell}

In \Cref{sec:NR-RFT} we introduced the pole-dominance approximation, in which $\cM_2$ is approximated by the leading order term in the Laurent expansion, $\cM_2^p$ of \Cref{eq:M2pole}. Upon NR reduction the latter is proportional to the two-body Green function in the LS approach, see Eq.~\eqref{eq:M2NR}. As noted in the main text, this approximation does not follow from a particular choice of power counting, and in general there can be significant nonpole contributions. These contributions incorporate important physical effects, absent in the LS approach, arising from $D\pi$ interactions away from the $D^*$ mass. In particular, nonpole terms are needed in order for $\cM_2$ to be unitary above the $D\pi$ threshold. In this regard, we stress again that, in the numerical results of \Cref{sec:numerical}, we use the full $\cM_2$ in the RFT calculations.

In this appendix we study the impact of nonpole terms on the matching between RFT and LS approaches. We write the two-body amplitude as
    \begin{align}
    \cM_2(q) = \cM_2^p(q) + \cB(q) \, ,
    \end{align}
where the nonpole or ``background'' contribution, $\cB(q)$, is a regular function below the $D\pi$ threshold, $\sigma(q) < (m_D + m_\pi)^2$. This term may be a low-order polynomial in $\sigma_q$, and we do not specify here its exact form.

Performing the NR reduction of the RFT equations as in the main text but with $\cM_2^p$ replaced by $\cM_2$ leads to a new form of the two-particle Green function in the LS equation,
    \begin{align}
    \label{eq:BNR}
    \cM_2(q) \approx \frac{\zeta^2}{2 m_{D}} \, \widetilde{\cG}(q; E) \,
    , \quad  \widetilde{\cG} = \cG + \cB_{\rm NR}\, .
    \end{align}
Here $\cG$ is defined in Eq.~\eqref{eq:M2NR} and $\cB_{\rm NR}$ is the non-relativistic approximation of $\cB(q)$. 
The function $\widetilde \cG$ appears in both the reduction of the ladder amplitude $\cD$ (see \Cref{sec:RFT-NR-ladder}) and the divergence-free remainder $m_{\rm df,3}$ (see \Cref{sec:mdf3reduce}).
It follows that one obtains the full LS equation, except that $\cG$ is replaced by $\widetilde \cG$, i.e.
    \begin{align}
    \widetilde{T}(p', p; E) = V^{\rm NR}(p',p) + \int_0^\Lambda \frac{dq \, q^2}{2\pi^2} \, V^{\rm NR}(p',q) \, \widetilde{\cG}(q; E) \, \widetilde{T}(q,p;E) \, ,
   \label{eq:mod-LS-1}
    \end{align}
    where
    \begin{equation}
    V^{\rm NR}(p',p) = V_{\pi,S}^{\rm NR}(p',p) + 2 \widetilde c_0\,,
    \end{equation}
with $V_{\pi,S}^{\rm NR}$ given in \Cref{eq:VpiSc}.
Here we are placing tildes on all quantities entering equations in which
$\tilde \cG$ replaces $\cG$. We note that the argument of \Cref{sec:mdf3reduce} implies that $\widetilde c_0$ is given in terms of $\widetilde\cK_3^E$ by \Cref{eq:matching-c0} (with tildes added). Here $\widetilde{\cK}_3^{E}$ is the value that would be obtained from a fit of the RFT formalism were one to fit lattice (or experimental) data including the nonpole term $\cB$ in $\cM_2$.

Our next step is to rewrite \Cref{eq:mod-LS-1} so that $\widetilde \cG$ is replaced by $\cG$. This can be accomplished by changing the potential to $\widetilde{V}$, which satisfies the integral equation
    \begin{align}
    \widetilde{V}(p',p) = V^{\rm NR}(p',p) + \int_0^\Lambda  \frac{dq \, q^2}{2\pi^2} \, V^{\rm NR}(p',q) \, \cB_{\rm NR}(q; E) \, \widetilde{V}(q,p) \, .
    \label{eq:Vtilde}
    \end{align}
It is then straightforward to show that $\widetilde T$ satisfies
    \begin{align}
    \label{eq:mod-LS-2}
    \widetilde{T}(p', p; E) = \widetilde{V}(p',p) + \int_0^\Lambda \frac{dq \, q^2}{2\pi^2} \, \widetilde{V}(p',q) \, \cG(q; E) \, \widetilde{T}(q,p;E) \,,
    \end{align}
so that $\widetilde T$ is a solution of the full LS equation with the modified potential. This will lead to additional contributions to the matching of parameters in the RFT and LS approaches.

So far, the analysis has made no approximations beyond NR reduction. We now assume that $\cB_{\rm NR}$ is small, so that the integral equation~\Cref{eq:Vtilde} can be expanded in a Born-like series. Keeping the first nontrivial term, we then have
    \begin{align}
    \widetilde{V}(p',p) \approx V^{\rm NR}(p',p) + \int_0^\Lambda  \frac{dq \, q^2}{2\pi^2} \, V^{\rm NR}(p',q) \, \cB_{\rm NR}(q; E) \, V^{\rm NR}(q,p) \, .
    \label{eq:Vtilde-b}
    \end{align}
The second term provides an additional contribution to the potential that depends both on the OPE potential $V_{\pi,S}$ and on $\widetilde{c}_0$, and leads to an additional non-trivial dependence of the interactions on off-shell momenta. It also acquires an implicit dependence on the total energy. Thus, it contributes to all orders in the chiral EFT expansion. Expanding in powers of momenta, and keeping the leading order term, gives a new contact potential, with the coupling constant becoming
    \begin{align}
    c_0 = \widetilde{c}_0 + \delta c_0 \, , \quad  \delta c_0 = \frac{1}{2} \int_0^\Lambda  \frac{dq \, q^2}{2\pi^2} \, V^{\rm NR}(0,q) \, \cB_{\rm NR}(q; E_{\rm th}) \, V^{\rm NR}(q,0) \, ,
    \label{eq:deltac0}
    \end{align}
with $E_{\rm th}=m_D+m_{D^*}$. Thus, within the discussed approximations, the short-range LS couplings obtained from matching to the RFT equations with the full $\cM_2$ are shifted by $\delta c_0$ compared to using only the pole term $\cM_2^p$ in the RFT approach.

We have carried out a numerical investigation of the approximate formula for $\delta c_0$, \Cref{eq:deltac0}. Using the parameters described in \Cref{sec:numerical}, i.e., for $\widetilde{\cK}_3^E = 4 \cdot 10^4$ and the resulting $\widetilde{c}_0 m_D^2 = -11.75$, we obtain,
    \begin{align}
    \delta c_0 \, m_D^2 \approx -4.66  \, .
    \end{align}
The non-relativistic approximation of $\cM_2(q)$, required to obtain $\cB_{\rm NR}$ in~\Cref{eq:BNR}, was taken by using $\sigma(q) \approx (E_{\rm NR} - m_{D^*})^2 - q^2 \, m_{D^*}/\mu$ in~\Cref{eq:qqstars,eq:2b-kmatrix}.

\begin{figure}[t!]
    \centering
    \includegraphics[width=0.98\textwidth]{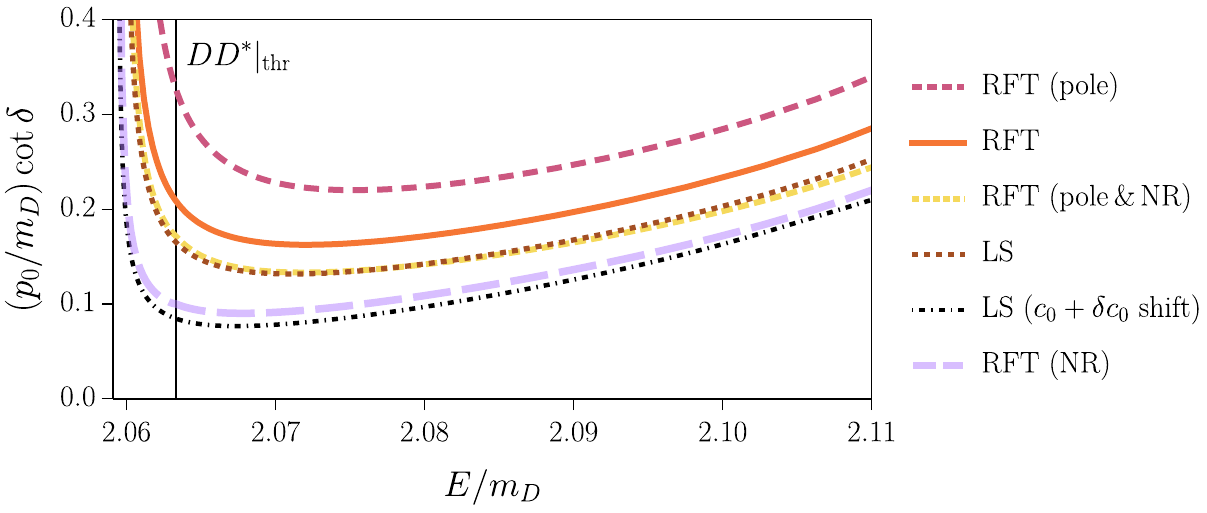}
    \caption{Comparison of $DD^*$ $s$-wave phase shifts obtained using the three-body RFT and LS approaches in various approximations. The parameters are as in \Cref{fig:phase-shifts}. See text for more details. For clarity, only results above the left-hand cut are shown.
    }
    \label{fig:compare-approximations}
\end{figure}

In~\Cref{fig:compare-approximations}, we show the impact of this shift on the LS result---this can be seen by comparing the original LS curve, labeled ``LS'', with that labeled ``LS ($c_0+\delta c_0$ shift)''. We see that it is significantly lowered. We also compare the shifted curve to various results obtained using the RFT formalism: the full RFT result (which is the same as that in \Cref{fig:phase-shifts,fig:phase-shifts-best-match}), the RFT result keeping only the pole term in $\cM_2$ [labeled ``RFT (pole)''], the RFT result keeping the full $\cM_2$ but making the NR approximation [``RFT (NR)''],\footnote{%
The NR approximation is made by replacing  all the building blocks of the amplitude ($\cM_2$, $G$, $\widetilde{\rho}$, $\cK_3$, integration Jacobian, on-shell momenta, etc.) by their NR approximants. Specifically, we use~\Cref{eq:NRmomenta} for the NR momenta,~\Cref{eq:GNR} for the OPE amplitude, and~\Cref{eq:KLNR,eq:roNR} in the expression for $m_{\rm df,3}$.
}
and the RFT result keeping only the pole term and making the NR approximation [``RFT (pole \& NR)'']. We observe that the latter result essentially agrees with the unshifted LS result, confirming the analytic results of \Cref{sec:NR-RFT}.\footnote{%
The small numerical deviation between two curves is explained by an inconsistent treatment of the reduced mass $\mu$ (which becomes $m_D/2$ at LO of the $\Delta$ expansion in ``RFT (NR)'' equations) in both approaches.}
The key result, however, is that the ``RFT (NR)'' curve lies close to the shifted LS curve, validating the analysis of this appendix and in particular the approximate result \Cref{eq:deltac0}. We also observe that the impact of the pole-dominance and NR approximations alone act in opposite directions, the former raising the RFT result while the latter lowering it.

In summary, we have shown that the matching of RFT formalism to the LS approach can, in the NR approximation, be extended to include the effects of nonpole terms in the $D\pi$ amplitude.

\bibliographystyle{JHEP} 
\bibliography{main}

\end{document}